\documentclass{emulateapj}
\newcommand{\vvec}{\mathbf{v}} \newcommand{\jvec}{\mathbf{j}}
\newcommand{\alphaMRI}{\alpha_{\rm MRI}}

 \newcommand{\Mdot}{\dot{M}}
\newcommand{\Msun}{M_{\odot}} \newcommand{\tff}{t_{\rm ff}}
\newcommand{\krho}{k_{\rho}} 
\newcommand{\tffs}{t_{\rm ff,s}}
\begin{document}
\title{Global Models for the Evolution of Embedded, Accreting
  Protostellar Disks} \author{Kaitlin M. Kratter and Christopher
  D. Matzner} \affil{Department of Astronomy and Astrophysics,
  University of Toronto, Toronto, ON M5S 3H4}

\author{Mark R. Krumholz\altaffilmark{1}} \affil{Department of
  Astrophysical Sciences, Princeton University, Princeton, NJ
  08544-1001}

\altaffiltext{1}{Hubble Fellow}

\begin{abstract} 
Most analytic work to date on protostellar disks has focused on those
in isolation from their environments. However, observations are now
beginning to probe the earliest, most embedded phases of star
formation, during which disks are rapidly accreting from their parent
cores and cannot be modeled in isolation. We present a simple,
one-zone model of protostellar accretion disks with high mass infall
rates. Our model combines a self-consistent calculation of disk
temperatures with an approximate treatment of angular momentum
transport via two mechanisms. We use this model to survey the
properties of protostellar disks across a wide range of stellar masses
and evolutionary times, and make predictions for disks' masses, sizes,
spiral structure, and fragmentation that will be directly testable by
future large-scale surveys of deeply embedded disks.  We define a
dimensionless accretion-rotation parameter which, in conjunction with
the disk's temperature, controls the disk evolution.  We track the
dominant mode of angular momentum transport, and demonstrate that for
stars with final masses greater than roughly one solar mass,
gravitational instabilities are the most important mechanism as most
of the mass accumulates.  We predict that binary formation through
disk fission, fragmentation of the disk into small objects, and spiral
arm strength all increase in importance to higher stellar masses.
\end{abstract}

 
\section{Introduction}\label{intro}

A young star system's visible T~Tauri or Herbig stage is preceded by a
deeply enshrouded phase of rapid accretion in which its character --
multiplicity, disk properties, and tendency to form planets -- is
first forged.  Although this embedded phase is likely the one during
which most accretion onto the star occurs, the properties of disks
during this period have received relatively little attention.  This
phase is difficult to model analytically because embedded disks are
subject to large perturbations in the form of rapid accretion of mass
and angular momentum, making local models and stability analyses
problematic. Due to the high obscuration characteristic of this phase,
disks are accessible primarily via radio and submillimeter
observations, and such techniques provide limited sensitivity and
angular resolution compared to what can be achieved for T~Tauri and
Herbig AE star disks using shorter wavelengths. Our knowledge of
massive protostellar disks is particularly limited by this problem,
since they do not have a significant unobscured phase, probably due to
the destructive effects of ionizing radiation.  These difficulties are
compounded by the fact that massive stars form more rarely, and
therefore tend to lie farther away.  Detections of rotation and infall
in a few systems hint at the presence of disks during the embedded
phase, but are only preliminary (see recent reviews by
\citealt{cesaroni06, cesaroni07a}, \citealt{beuther06b}, and
\citealt{beuther07a}).

While our knowledge of the embedded phase today is limited, it will
soon come into sharp focus as new instruments such as the Expanded
Very Large Array (EVLA) and Atacama Large Millimeter Array (ALMA)
become operational. In order to predict what these telescopes will
discover about the formation of stars across a very broad mass range,
$\sim 1-120 \Msun$, we present evolutionary models of star-disk
systems reacting to infall at very high rates.  We concentrate our
efforts on the physical processes that control disk evolution, such as
the torque from a turbulent infall, the reprocessing of starlight by
the infall envelope, and the character of the self-gravitational
instabilities.  The disk itself we model with a highly simplified,
single-zone treatment.  Although multidimensional simulations provide
a much more detailed view of disk formation and evolution during the
embedded phase \citep[e.g.][]{2004A&A...414..633G, Krum2007a}, their
high computational cost and the limited range of physical processes
they include mean that these simulations can explore only small
regions of parameter space. They cannot easily make predictions across
a broad range of stellar mass scales and evolutionary times. In
contrast, our semi-analytic approach permits us to incorporate more
physical effects and explore the consequences of environmental
parameters more rapidly, and in a more systematic way. This serves two
complementary ends: the theoretical goal of understanding angular
momentum transport and fragmentation in the embedded phase, and the
observational goal of making concrete predictions about the properties
of young, massive disks.

Although we include a range of physical processes in our models;
the most important in driving the evolution of embedded disks
in our calculations is self-gravity. Self-gravity plays a central role
in mediating angular momentum transport and triggering fragmentation
into a binary or multiple system.  Its importance in star formation
has long been recognized \citep{1984MNRAS.206..197L}, and our study is
preceded by evolutionary calculations which incorporate accretion and
self-gravity into one-dimensional
\citep{1987MNRAS.225..607L,1990ApJ...358..515L,1994ApJ...421..640N,
  1995ApJ...445..330N,2005A&A...442..703H} and two-dimensional
\citep{2005ApJ...633L.137V,Vorobyov06} simulations.  Although lower in
resolution, our approach is distinguished from these works in several
ways:
\begin{enumerate}
\item[$i$.] In contrast to all one-dimensional calculations to date,
  we account for the dependence of gravitational torques on the
  disk-to-total mass ratio in addition to Toomre's instability
  parameter.
\item[$ii$.] We consider the possibility that disks will fragment if
  they become sufficiently unstable.
\item[$iii$.] We consider fluctuations of the vector angular momentum
  in the infall due to realistic turbulence in the collapsing cloud
  core.
\item[$iv$.] We employ a realistic model for the irradiation of the
  disk midplane, in which starlight is reprocessed at the inner wall
  of an outflow cavity while inflow is occurring.
\item[$v$.]  We survey the conditions of intermediate-mass and massive
  star formation, rather than focusing exclusively on conditions in
  nearby low-mass star forming regions such as Taurus.
\end{enumerate}
The first of these is important for all protostellar disks, since the
disk mass is never negligible when the Toomre parameter is small.
Features ($ii$) and ($iii$) are of primary importance in the formation
of massive stars, which accrete from strongly turbulent regions
\citep{1992ApJ...396..631M,1997ApJ...476..750M} and which are likely
to undergo disk fragmentation (\citealt{KM06}, hereafter KM06).
Irradiation is most important for low-mass stars, whose disks
it strongly stabilizes
\citep{ML2005}, but it remains significant in massive star formation
as well.

In \S \ref{basics} we outline our model for the infall rate of matter
and angular momentum.  We develop a model for disk accretion and
fragmentation in \S \ref{diskdynamics}. In \S \ref{expects} we define
the key environmental variables that control protostellar disk
evolution and sketch a qualitative evolutionary sequence based on
their fiducial values. In \S \ref{analysis} we present the results for
our fiducial case, and explore the effect of varying our
parameters. In \S \ref{observables} we discuss the observable
predictions that our model makes, and finally in \S~\ref{conclusions}
we summarize our main results.

\section{Infall onto Disks}\label{basics}

Since we are interested in the behavior of a disk that is subject to
strong perturbations from its environment, we begin building our model
by constructing a prescription for the infall of matter and angular
momentum onto a disk. This accretion comes from a background ``core''
\citep{Shu1977, 1997ApJ...476..750M,MT2003}, whose properties and
interaction with a disk we discuss in this section.

\subsection{Star Formation By Core Collapse} \label{sfmodel}

We model the accretion of mass and angular momentum using the
two-component core model of \citet{MT2003}, which is a generalization
of the TNT (thermal plus non-thermal) model of
\cite{1992ApJ...396..631M}. In this model, which we summarize here for
convenience, the density distribution within a core follows a
two-component power law distribution
\begin{equation}
\label{rho_core}
\rho = \rho_s \left(\frac{R_{\rm c}}{r}\right)^{-k_\rho} +
\frac{c_{\rm s, c}^2}{2\pi G r^2},
\end{equation}
where $R_{\rm c}$ is the outer radius of the core, $c_{\rm s,c}$ is
the thermal sound speed within it (assumed to be constant), and
$\rho_s$ is the density at the core's surface. We follow \cite{MT2003}
in adopting $k_\rho= 1.5$ as the fiducial value of the
turbulence-supported density index. Physically, the first term
describes an envelope supported primarily by turbulent motions, while
the second describes a thermally-supported region at its center. A
model of this sort is fully specified in terms of the three
parameters: the core mass
\begin{equation}
M_{\rm c} = \frac{4\pi}{3-k_\rho} \rho_s R_{\rm c}^3 + 2
\frac{c_{s,c}^2 R_{\rm c}}{G},
\end{equation}
surface density
\begin{equation}
\Sigma_{\rm c} = \frac{M_{\rm c}}{\pi R_{\rm c}^2},
\end{equation}
and temperature
\begin{equation}
T_{\rm c} = \frac{m}{k_B} c_{s,c}^2,
\end{equation}
where $m=3.9\times 10^{-24}$\,g is the mean particle mass in a gas of
molecular hydrogen and helium mixed in the standard cosmic
abundance. Observed regions of star formation contain cores with
masses $\sim 1-100$ $\Msun$, surface densities $\Sigma_{\rm c} \approx
0.03-1$ g cm$^{-2}$, and temperatures of 10 to 50 K
\citep{2001ApJ...559..307J,1997ApJ...476..730P}.

The core is taken to be in approximate hydrostatic balance initially,
and this condition specifies the required level of  turbulent
support. The non-thermal velocity dispersion in the shell at radius
$r$ is
\begin{equation} \label{sigma_core}
\sigma(r)^2 = {2\pi\over 3\phi_B(\krho-1)} {G M(r)\over r} - c_{\rm
  s,c}^2
\end{equation}
where $M(r)$ is the mass at radii of $r$ or less and $\phi_B\simeq
2.8$ approximately accounts for the magnetic contribution to the total
pressure. Except when $M(r)\ll 1$ $\Msun$ or $\Sigma_c <0.1$ g
cm$^{-2}$, the first term is much larger than the second, so that the
velocity dispersion is primarily non-thermal.

Core collapse commences at time zero, and a mass shell initially at
radius $r$ falls onto the disk in a time comparable to the free fall
time evaluated at its initial density, $\tff(r) = [3\pi /
  32G\rho(r)]^{1/2}$. In practice, we use the \cite{MT2003} accretion
rate approximation
 to determine $\Mdot$ on to the star-disk system as a
function of the total core mass and the current amount of mass that
has accreted:

\begin{equation}\label{MTacc}
 \Mdot_{\rm in} \approx \left(\frac{\phi_* M_{*f}}{\tffs}\right)
 \left[\left(\frac{M_{*d}}{M_{*f}}\right)^{2q} + \left(\frac{\phi_{*\rm
       th}}{\phi_{*\rm nth}}\right)^2 \left(\frac{\varepsilon M_{\rm
       th}}{M_{*f}}\right)^{2q}\right]^{1/2},
 \end{equation}
where $M_{*f}$ is the final disk plus stellar mass, $M_{*d}$ is the
current disk plus stellar mass, $\tffs$ is the free-fall time
evaluated at the core surface (i.e.\ at $\rho=\rho(R_c)$),
\begin{eqnarray}
q & = & \frac{3(2-k_\rho)}{2(3-k_\rho)}, \\ M_{\rm th}
&= &10^{-3.1}
\left({\frac{T}{20\,\rm K}}\right)^3 \left(\frac{30 \varepsilon
  \Msun}{M_{*f}}\right)^{1/2} \Sigma_{c,0}^{-3/2} \Msun,
\end{eqnarray}
$\Sigma_{\rm c,0}=\Sigma_{\rm c}/(\mbox{g cm}^{-2})$, and $\phi_*$,
$\phi_{*{\rm nth}}$, and $\phi_{*\rm th}$ are constants of order unity
that depend on the polytropic index and magnetic field strength.
The efficiency factor
\begin{equation} \label{epsilon}
\varepsilon = {M_{*f}\over M_c}
\end{equation} 
represents the fraction of the core mass that lands on the star-disk
system rather than being blown out by protostellar outflows.  We again
follow \cite{MT2003} in adopting $\varepsilon =0.5$, a value typical
of low-mass star formation \citep{MM2000}.

\subsection{Angular Momentum of the Infalling Material}\label{j_acc}

Equation (\ref{MTacc}) gives the mass infall rate $\dot{M}_{\rm in}(t)$
from the core as a function of time. The second component of our core
model is to specify the corresponding rate of angular momentum infall
$\mathbf{\dot{J}}_{\rm in}(t)$. We compute
this in several steps. First, we approximate the vector specific angular
momentum $\jvec(r)$ averaged over a shell of material at radius
$r$ as described below. Then we compute $M(t)\equiv \int_0^t
\dot{M}_{\rm in}(t')/\epsilon\, dt'$, the total mass from the core that
has either fallen onto the star-disk system or been ejected at time
$t$. From the core density profile (equation \ref{rho_core}) we also
compute $M(r)\equiv \int_0^r 4\pi r'^2 \rho(r')\, dr'$, the mass
of the initial core enclosed within radius $r$. Assuming the core
accretes inside-out, we set $M(r)=M(t)$ and solve for $r(t)$, which
gives the initial radius $r$ of the shell of mass that arrives at the
star-disk system at time $t$. Assuming that the specific angular
momentum of the gas does not change before it reaches the disk, the
angular momentum accretion rate is then simply given by
$\mathbf{\dot{J}}_{\rm in}(t) = \dot{M}_{\rm in} (t) \jvec(r(t))$.

The remaining step is to specify how we estimate $\jvec(r)$. 
Star forming cores are often modeled as solid body rotators
characterized by the ratio $\beta$ of rotational to gravitational
energy,
but we adopt a more realistic model in which turbulent fluctuations
affect the infalling gas.  Following \cite{2000ApJ...543..822B},
\cite{2004ApJ...600..769F}, \cite{ML2005} and KM06, we assume that the
observed angular momenta of cores \citep{1993ApJ...406..528G} can be
modeled using an idealized turbulent velocity field.
Using the method of \cite{1995ApJ...448..226D}, we generate a
numerical realization of an isotropic, homogeneous, Gaussian random
velocity field $\vvec(\mathbf{r})$.
We require that the power spectrum $P(k)$ of this turbulent field
reproduce
the scalings required by turbulent support against gravity:
$\sigma(r)^2 \propto GM(r)/r \propto r^{2-k_\rho}$ at large radii, so
that $\sigma(r) \propto r^{1/4}$ for $\krho = 3/2$.  Parceval's
theorem or dimensional analysis then require $P(k)\propto k^{-3/2}$.

We note that numerical simulations of supersonic turbulence
consistently show the steeper spectral index $-2$
\citep{1992PhRvL..68.3156P} which is understood as the spectrum of an
individual shock and as the exact limit of Burgers
turbulence. \cite{2007ApJ...659.1394M} and \cite{2007ApJ...662..395N}
have shown that a shallower index is expected when turbulence is
driven by protostellar outflows, however, and our chosen power
spectrum is consistent with the line width-size relation for massive
cores \citep[e.g.,][]{1995ApJ...446..665C,1997ApJ...476..730P}.  Our
homogeneous velocity field is surely an idealization, but not a grave
one.

After scaling our numerical domain to match the core radius $R_c$, we
normalize $\vvec$ such that the one-dimensional velocity dispersion of
a spherical shell with this radius equals $\sigma(r)$ defined in
equation (\ref{sigma_core}).  In practice we fit $R_c$ within a
$256^3$ section of a $1024^3$ grid of velocities, because periodicity
causes artifacts on scales larger than about 1/4 of the box size.
From this field we calculate the specific angular momentum $\jvec =
\mathbf{r} \times \mathbf{v}$ at every point and the mean specific
angular momentum $\jvec(r)$ in a shell at radius $r$.
Note that KM06 calculate the expected magnitude and dispersion of
$\jvec(r)$ for velocity fields of precisely this type; our results
agree with their predictions to about 50\%, which is within the
scatter they predict.

\section{Dynamics of the disk}\label{diskdynamics}

\subsection{Approach to disk evolution} \label{Approach}

Given the rate at which mass and angular momentum accrete, we must
calculate the reaction of the disk. At any given time, our
star-plus-disk system is characterized by the disk mass $M_d$, the
central star mass $M_*$, and the total angular momentum content of the
disk $\mathbf{J}_d$. Given these quantities, and the rates of mass and
angular momentum infall $\dot{M}_{\rm in}$ and $\dot{\mathbf J}_{\rm
  in}$, we wish to compute the time rate of changes $\dot{M}_d$,
$\dot{M}_*$, and $\mathbf{\dot{J}}_d$. 

Using the separation between the thermal, orbital, and accretion
timescales, we assume our disks are in a thermal steady state and
draining at a rate determined by their current global parameters.  We
shall later refer to this as the assumption of thermal and mechanical
equilibrium. 

In \S~\ref{Mdotstar}, we estimate the disk accretion rate onto the central star due to various angular momentum transport mechanisms. In \S~\ref{diskheat}
we discuss thermal equilibrium in the disk, which together with the
aforementioned condition of mechanical equilibrium allows us to self-consistently
compute the accretion rate from the disk to the star $\dot{M}_*$. In
\S~\ref{angmom} we describe the corresponding angular momentum
evolution $\dot{\mathbf J}_d$. Finally, in \S~\ref{fragmentation} and
\S~\ref{binaries}, we discuss our prescriptions for disk fragmentation
and binary formation.

It is helpful before proceeding to define two dimensionless parameters
that characterize the strength of the disk's self-gravity. These are
the disk-to-total mass ratio
\begin{equation}
\mu = {M_d\over M_d+M_*}
\end{equation}
and \citeauthor{Toom1964}'s (\citeyear{Toom1964}) instability
parameter
\begin{equation}\label{Qtoomre} Q={c_s \kappa\over \pi G \Sigma_d },
 \end{equation} 
where $\kappa$ is the radial epicyclic frequency, $c_s$ speed of
density waves, and $\Sigma_d$ is the disk's mass surface density.  In
practice we evaluate $Q$ using $\kappa \rightarrow \Omega =({GM_{\rm
    tot} / R_d^3})^{1/2}$, the total orbital frequency, since the
difference between them is only marginally significant even when the
disk mass is quite large. Here $R_d = j_{d}^2 / {G M_{\rm tot}}$. We
also approximate $c_s$ using the isothermal sound speed.  To
characterize gravitational instability, we use the {\em minimum} value
of $Q$ -- the value at $R_d$, the outer boundary of our active disk.
In this evaluation we assume a profile $\Sigma_d \propto r^{-1}$
(a choice we justify in \S~\ref{angmom})
so that $\Sigma_d(R_d) = M_d/(2\pi R_d^2)$, and we evaluate $c_s$ and
$\Omega$ at the edge of the disk.

We base our models on the fundamental assumption that the
self-gravitational behavior of an accretion disk depends primarily on
the structural parameters $\mu$ and $Q$ -- so that its evolution is
controlled by heating and cooling (\S~\ref{diskheat}), which alter
$Q$, and accretion onto and through the disk, which alters $\mu$.
This approach permits us to treat the disk's mechanical and thermal
properties separately, before combining them into a model for its
evolution.  This division also guides our use of published work, since it implies
that simulations with adiabatic equations of state and those with an
imposed cooling rate may be combined into a mechanical model for disk
evolution, which we may then use to model irradiated protostellar
disks.  Finally, it prompts us to treat the onset of disk
fragmentation and disk fission as boundaries in the space of $\mu$ and
$Q$, rather than in terms of a critical cooling rate (which is the
natural and conventional description for simulations that include
cooling but not irradiation).  In \S~\ref{fragmentation} we argue that
these descriptions are effectively equivalent.

Models based on this assumption are guaranteed to be somewhat
approximate, because a disk's mechanical evolution must, at some
level, reflect additional parameters: its dimensionality, its equation
of state, the specifics of its heating and cooling processes, and the
magnitude of external perturbations (like tides), to name a few.
However we expect our results to be valid, both because we believe
that $\mu$ and $Q$ are indeed the most significant parameters for
gravitational instability, and because our model is calibrated to
realistic numerical models.  Additional simulations will be required to
test this approach.

\subsection{Angular Momentum Transport and Disk Accretion}\label{Mdotstar}

A key element of our model is a prescription for angular
momentum transport and the rate $\Mdot_*$ at which matter accretes onto the
central star -- or more specifically, the dimensionless rate
$\Mdot_*/(M_d \Omega)$.  In practice we first construct a model for an
effective \citeauthor{SS1973} $\alpha$ parameter, which we define
through the steady-state relation
\begin{equation}\label{mdotalpha}
\dot{M}_{*} = \frac{3 \alpha c_s^3}{GQ}
\end{equation}
so that
\begin{equation}\label{mdotloc}
{\Mdot_{*}\over M_d \Omega} = \frac{3\alpha Q^2}{8}\mu^2,
\end{equation}
where the factor of of $3/8$ comes from the assumption that the disk
surface density falls as $r^{-1}$.
We do not mean to imply by this that transport induced by
gravitational instability is purely local \citep{BP99}, although this
does appear to be the case for sufficiently thin and light disks
\citep{Gam2001,LodRi05}.

We divide $\alpha$ into two contributions: $\alpha_{\rm MRI}$, due to
the magnetorotational instability (MRI), and $\alpha_{\rm GI}$, due to
gravitational instability.  In keeping with the strategy described in
\S~\ref{Approach}, we consider $\alpha_{\rm GI}$ to be a pure function
of $\mu$ and $Q$.  We combine it and the MRI contribution linearly:
\begin{equation} \label{alphaTotal}
\alpha = {\alpha_{\rm GI} + \alpha_{MRI}}.
\end{equation}
We create our model for $\alpha_{GI}(Q,\mu)$ using results from the
simulations of \cite{1996ApJ...456..279L}, \cite{2003MNRAS.339.1025R},
\cite{Lod04}, \cite{LodRi05} and \cite{Gam2001}.  We adopt a constant
value for $\alpha_{MRI}$, as discussed below.

\subsubsection{Overview of Simulations} \label{lit}
The three sets of simulations span a large fraction of our parameter
space in $Q$ and $\mu$.  The global disk simulations of
\cite{1996ApJ...456..279L} explore $Q>1$ and non-negligible values of
$\mu$ using a two-dimensional hydrodynamic, self-gravity code; they
supress local fragmentation by imposing an adiabatic equation of
state. The simulations of \cite{Gam2001} represent the limit $\mu
\rightarrow 0$, for values of $Q$ which approach unity from above, and
are most directly applicable to quasar disks.
\citeauthor{Gam2001} imposes cooling with a fixed
cooling time, $\tau_c$, which is proportional to the orbital time.  He
finds a regime of steady gravity-induced turbulence, for disks that
cool over many orbits.  If $\tau_c$ is too short ($< 3\Omega^{-1}$),
however, the disk fragments as $Q$ drops below unity.  The disk
viscosity is highest at the boundary of fragmentation.  Angular
momentum transport in this regime is quite local, with an effective
value of $\alpha$ that is inversely proportional to the cooling
rate. Our third source is the global SPH simulations of
\cite{2003MNRAS.339.1025R}, \cite{Lod04}, and \cite{LodRi05},
in which a cooling time $\propto \Omega^{-1}$ is imposed
locally; these cover the entire parameter space in $\mu$.  In these
simulations
$Q$ is initially 2, but it descends towards unity. Here again, the
disk fragments if $\Omega \tau_c$ is too small, although the critical
value of this parameter is different than
\citeauthor{Gam2001} found.

\subsubsection{Accretion Model}
\begin{figure}
\plotone{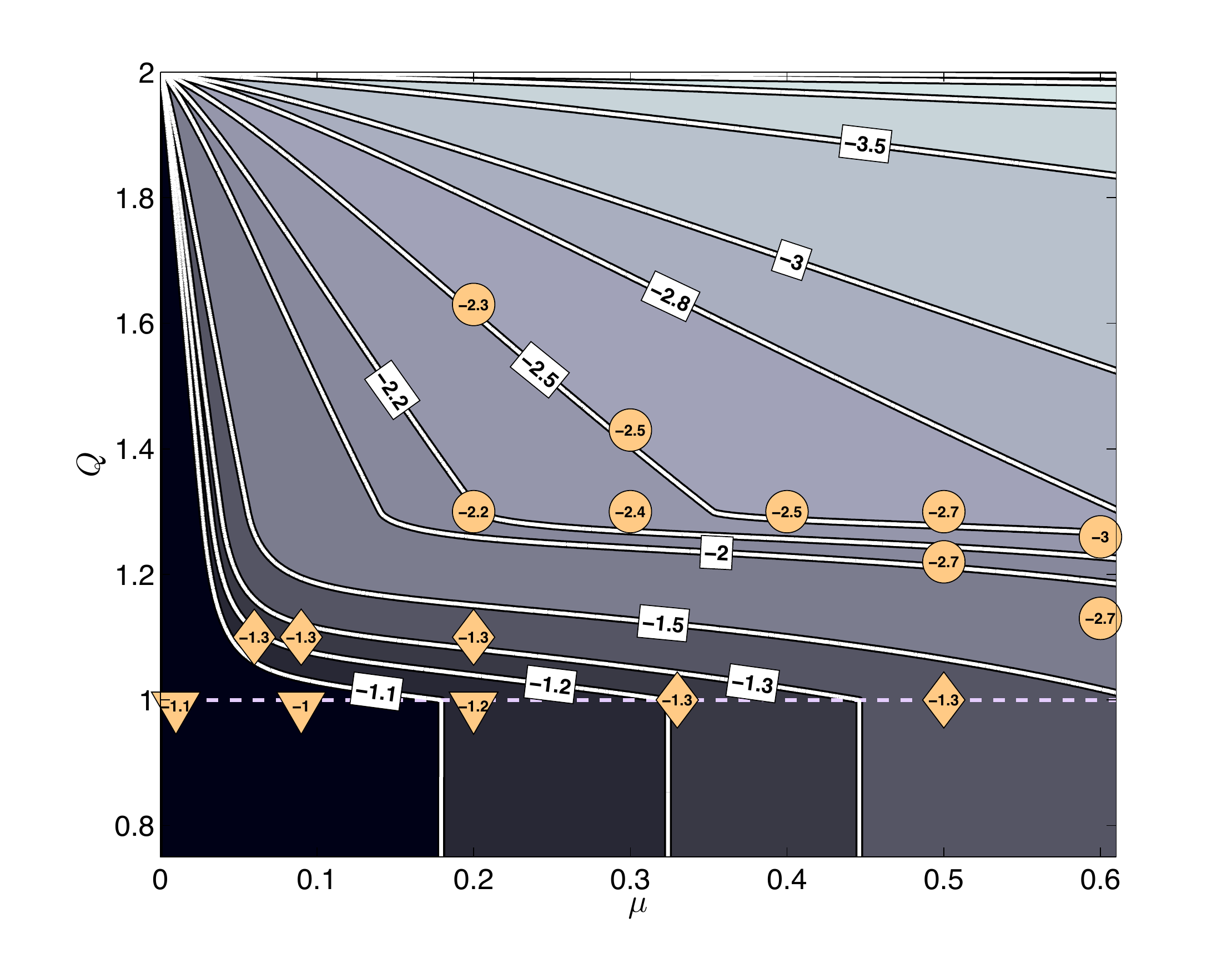}
\caption{{Contours of the viscosity parameter log($\alpha_{\rm GI}$)
    due to gravitational instabilities (eq.~\ref{alphaGI}); white
    squares are contour labels. Results from numerical simulations are
    marked with circles, diamonds, and triangles. Circles show
    simulations with adiabatic equations of state
    \citep{1996ApJ...456..279L}, diamonds show simulations with an
    imposed cooling rate that reach steady state \citep{Lod04,
    LodRi05}, and triangles show the maximum $\alpha_{\rm GI}$
    achieved in simulations with imposed cooling that probe the
    fragmentation boundary \citep{Gam2001, 2003MNRAS.339.1025R, Lod04,
    LodRi05}. Note that the point at $\mu=0$ corresponds to the purely
    local simulation of \citet{Gam2001}. The $Q=1$ boundary is marked
    with a dashed line.}}
\label{datacomp}
\end{figure}

To derive a relatively simple analytic fit to the simulation data, we
must extract a characteristic $\alpha_{\rm GI}, Q,$ and $\mu$ from the
simulations listed above.  Because the numerical approaches are
varied, we are unable to use the same method for each. The values are
derived as follows for each type of simulation:
\begin{enumerate}
\item {
We estimate
$\alpha_{\rm GI}$ from \cite{1996ApJ...456..279L}
using their equations 24-26; $Q$ and $\mu$ are given.}
\item {
Values of
 $\alpha$ from \cite{2003MNRAS.339.1025R}, \cite{Lod04},
\cite{LodRi05} are
taken directly from plots, when available.  Because $\alpha$ varies
with radius,
we take an approximate value from the outer region of their disk
before the density begins to fall off steeply. When plots are
not available, we use the critical value of $\tau_c$ to calculate
$\alpha_{\rm GI}$ at the fragmentation boundary, which we take to be
$Q=1$ (see \S~\ref{fragmentation}). Again, $\mu$ is given.
}
\item{\cite{Gam2001} provides one
value of $\alpha_{\rm GI}$ at $Q=1$ for a disk with $\mu\rightarrow0$,
which we adopt.}
\end{enumerate}
These values of $\alpha_{\rm GI}$ are shown in Figure \ref{datacomp}
according to the estimated values of $\mu$ and $Q$ that accompany each
of them.  We treat them as a data set to be fit within our analytical
model for $\alpha_{\rm GI}$, which is displayed in the underlying contours
in that figure.  Imposing the realistic condition that $\alpha_{\rm
  GI}$ is continuous and equals zero for $Q>2$ (when the gravitational
instability should shut off as suggested by \cite{2006MNRAS.365.1007G}), we find that two components are
required:
\begin{equation}\label{alphaGI}
\alpha_{GI} = \left(\alpha_{\rm short}^2 + \alpha_{\rm
  long}^2\right)^{1/2}
\end{equation} 
where
\begin{equation} 
\alpha_{\rm short} = \max\left[0.14\left(\frac{1.3^2}{Q^2}
  -1\right)(1-\mu)^{1.15},0\right]
\label{alphaLoc} 
\end{equation} 
and
\begin{equation} 
 \alpha_{\rm long} = \max\left[{1.4\times 10^{-3}(2-Q)\over\mu^{5/4}
     Q^{1/2}},0\right].
\label{alphaglob}
\end{equation}

In fact we apply equation (\ref{alphaGI}) only to the region $Q>1$.
Because we expect the gravitational torque to saturate when
fragmentation occurs, we hold $\alpha$ constant, for a given $\mu$,
when $Q<1$; this amounts to replacing $Q\rightarrow\max(Q,1)$ in the
above equations.  This has no practical consequences for our
calculations, however, since our treatment of fragmentation
(\S~\ref{fragmentation}) prevents our disks from sampling values of
$Q$ much below unity.

Our nomenclature in equation (\ref{alphaGI}) reflects our
interpretation.  The ``short'' component $\alpha_{\rm short}$
dominates for $Q\lesssim 1.3$, hence for relatively thin disks. We
think of it as arising from modes with relatively high spatial
wavenumbers and short wavelengths \citep{Lod04,LodRi05}.  Note that
its functional form resembles the model of \cite{1990ApJ...358..515L}
(their eq.~16) modified by a mild $\mu$ dependence, which is
comparable to the scale height dependence derived in equation (2.5) of
\cite{1987MNRAS.225..607L}.

The ``long'' component $\alpha_{\rm long}$ is important in thicker
disks whose instabilities are likely to be dominated by loosely wound,
$m=2$ spiral patterns.  We require it because we include the adiabatic
simulations of \cite{1996ApJ...456..279L}, which sample higher values
of $Q$ because they cannot cool.  (Indeed, $Q$ rises during these
simulations.) Our fundamental assumption (\S~\ref{Approach}) leads us
to incorporate these results into a single model for $\alpha(\mu,Q)$,
despite the difference in thermal physics.  Future simulations can
test this assumption by imposing heating (via irradiation, say) as
well as cooling: our model implies that the derived $\alpha_{\rm GI}$
will be comparable to \citeauthor{1996ApJ...456..279L}'s, when $Q$ and
$\mu$ take similar values.  
While simulations such as \cite{2006ApJ...651..517B} and \cite{2007arXiv0706.4046C} are making dramatic process towards accurately modelling heating, cooling, and irradiation, a  wider parameter space is necessary for comparison.
We note that \cite{SellCarl84}
and \cite{2006MNRAS.365.1007G} also find non-axisymmetric
instabilities for massive disks with $Q$ in the range $1.3-2$.

As shown in Figure \ref{datacomp} our model for $\alpha_{\rm GI}$
agrees reasonably well with data from the simulations, though we fail
to fit a couple points at very high $\mu$ and low $Q$. Note that $\alpha_{\rm GI}$ for these points from  \cite{1996ApJ...456..279L} are uncertain themselves.

It is important to bear in mind that our accretion model is only a
rough representation of the numerical results on which it is based,
and that it can be improved as more simulations become available.
For instance, we place no stock in the weak divergence of $\alpha_{\rm
  long}$ as $\mu\rightarrow0$: this feature is a product of our fit to
numerical results at larger $\mu$, and it would be an unwise
extrapolation to use our model for disks with very low $\mu$ and
moderate $Q$.  It does not affect our results, as our disks do not
sample this regime.

Finally, we assume the disk is sufficiently ionized to support
magnetic turbulence, and we represent the MRI with the constant value
$\alphaMRI = 10^{-2}$.  The typical value of $\alphaMRI$ is rather
uncertain; see \cite{2007arXiv0705.0352P} for a synthesis of recent
work, and \cite{2005A&A...442..703H} for a recent consideration of
observational constraints in low-mass protostellar disks.
Gravitational torques exceed those from the MRI for much of the
accretion phase. We discuss the influence of $\alphaMRI$ on our
results in \S\,\ref{MRI}

\begin{figure}
\begin{center}
\plotone{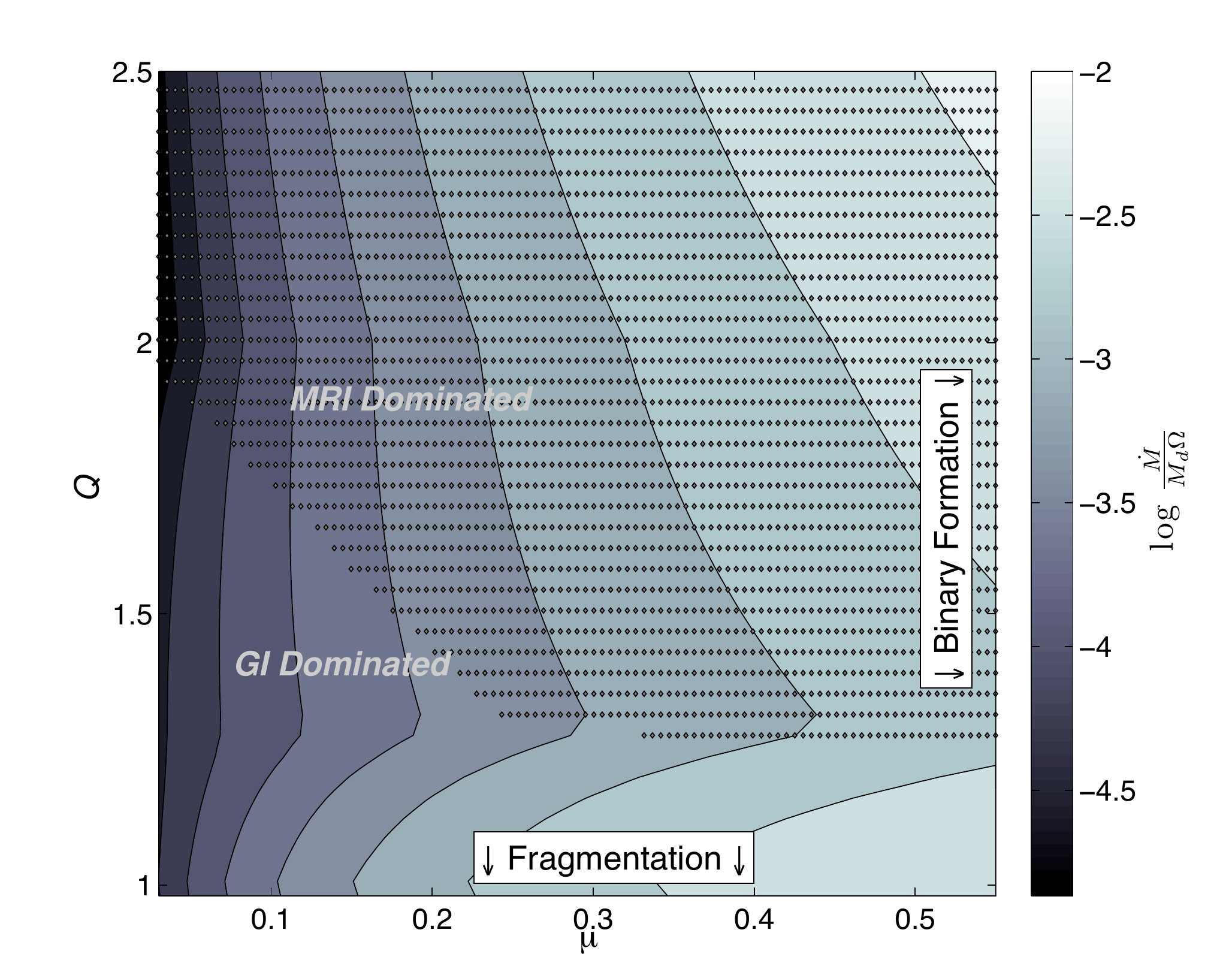}
\caption{{Contours of the dimensionless accretion rate from the disk onto the star ($\Mdot_*/ M_d
    \Omega$) from all transport components of our model.  The
    lowest contour level is $10^{-4.8}$, and subsequent contours increase by 0.3 dex. The effect of each transport mechanism is
    apparent in the curvature of the contours. At $Q<1.3$ the
    horizontal ``tongue" outlines the region in which short wavelength
    instability dominates accretion. The more vertical slope of the
    contours at lower $\mu$ and $Q>1.3$ shows the dominance of the
    long wavelength instability. The MRI causes a mild kink in the contours across the $Q=2$ boundary and is more dominant at higher disk
    masses due to our assumption of a constant $\alpha$: equation
    [\ref{mdotloc}] illustrates that a constant $\alpha$ will cause
    higher accretion rates at higher values of $\mu$.}}
\label{mdotcontours}
\end{center}
\end{figure}

Figure \ref{mdotcontours} illustrates our model for the dimensionless
accretion rate $\Mdot_*/(M_d\Omega)$ as a function of $Q$ and $\mu$.
We draw attention to several key features of the plot. First, note
that at low $Q$ there is a tongue-like feature that increases in
intensity with increasing disk mass. This is due to the strong
dependence of $\alpha_{\rm short}$ on both $Q$ and $\mu$
At higher values of $Q$ and lower values of $\mu$ the contours steepen
due to the weak divergence of $\alpha_{\rm long}$ as $\mu\rightarrow0$, which is probably not
physical. The curvature towards higher $Q$ and $\mu$ shows the dominance of the MRI
for $Q>2$. The fact that the dimensionless accretion rate takes
numerical values 
up to $10^{-2.4}$, with a typical values $\sim 10^{-3.5}$, 
implies that massive
disks drain on timescales ranging from $\sim 40$ to a few thousand 
orbits, with five hundred orbits being typical.

\subsection{Disk Thermal Equilibrium} \label{diskheat}
We have now specified the rate at which the disk accretes onto a
central star as a function of $Q$ and $\mu$. However, this does not
fully specify the accretion rate, because while $\mu$ may be directly
computed from our ``primitive'' variables $M_d$, $M_*$, and
$\mathbf{J}_d$, the Toomre stability parameter $Q$ cannot be, because
it depends on the sound speed $c_s$ and thus the temperature within
the disk. We can determine this by requiring that the disk be in
thermal equilibrium.

To compute the disk's thermal state, we follow the approach of KM06,
in which disks are heated by a combination of stellar irradiation and
viscous dissipation due to accretion. In equilibrium, the disk
midplane temperature satisfies the approximate relation
\begin{equation}
\label{Tdisk}
\sigma T^4 = \left(\frac{8}{3}\tau_R + {1\over4\tau_P}\right)F_v +
F_{\rm irr},
\end{equation}
where $F_v$ is the rate of viscious dissipation per unit area in the
disk, $F_{\rm irr}$ is the flux of starlight (whether direct or
reprocessed) onto the disk surface, and $\tau_{R,P} =
\kappa_{R,P}\Sigma/2$ are the Rosseland and Planck optical depths to
the midplane.  The viscous dissipation per unit area is
\begin{equation}
F_v = \frac{3\Mdot \Omega}{8\pi},
\end{equation}
and we compute the opacities using the \cite{2003A&A...410..611S}
model for $\kappa_{R,P}(T)$: in particular, we use their
homogeneous-aggregate dust model with normal silicates, calculated at
the appropriate density.

Low-mass stars' luminosities are accretion-dominated in the main
accretion phase, but those above about 15 $\Msun$ reach the zero-age
main sequence (ZAMS) while still accreting. To include both accretion
luminosity and other sources in our calculation of $F_{\rm irr}$, we
use the protostellar evolution code of \citet{krumholz07c}, based on
the \citet{MT2003} protostellar evolution model, to compute the
luminosity $L_*$ of the central star as a function of its accretion
history. The model includes contributions to the protostellar
luminosity from accretion on the stellar surface, Kelvin-Helmholtz
contraction, and, once the temperature rises high enough, deuterium
and then hydrogen burning.

During the infall, dust within the infall envelope reprocesses
starlight and casts it down on the disk.  By performing ray tracing
within an inflow envelope with a central outflow cavity, \cite{ML2005}
determine the fraction of light received by the disk assuming the
infall envelope is optically thick to the stellar radiation, and
optically thin to its own IR re-radiation: they find
\begin{equation}\label{Firr-in-accretion}
F_{\rm irr} = {0.1\over\varepsilon^{0.35}} {L_*\over 4\pi R_d^2}.
\end{equation} 
The weak dependence on the accretion efficiency $\varepsilon$ arises
from a picture in which the outflow clears a fraction $1-\varepsilon$
of the core, so that infall streamlines originate from regions
separated from the axis by angles $\theta$ such that $\cos \theta >
\varepsilon$.  Recently, \cite{2005ApJ...626..953R} have observed an
outflow near an O-type protostar with an opening angle of
approximately $25^\circ$; this is in reasonable agreement with the
model chosen here, since infall occurs at wider angles.

Once the core has accreted entirely and the envelope can no longer
re-process starlight, we make an (unrealistically) abrupt switch to a
model in which $\Mdot_{\rm in}=0$.  The star continues to acquire mass
from the disk, which represents a non-negligible reservoir.  From this
point on we calculate $F_{\rm irr}$ in the manner of
\cite{1997ApJ...490..368C}.  We first identify the fraction of $L_*$
intercepted by the surface which is optically thick to stellar
photons, assuming for this purpose that $H\propto R^{9/7}$ and that
the dust density is a Gaussian, of scale height $H$, in the height
above the midplane.  We also calculate the equilibrium temperature of
dust in this reprocessing layer.  We then calculate $F_{\rm irr}$ as
that fraction of the reprocessed radiation which is reabsorbed by the
disk, allowing for the possibility that the disk will be optically
thin at the relevant wavelengths. We find the reprocessing height is
slightly larger than a scale height (1.5\,$H$ being typical); higher
values are typical of more massive disks, which are more opaque.

Though negligible during the accretion phase, we also include a
background radiation field due to the cloud
(modeled as an optically thin dust layer)
and the cosmic microwave background.  This prevents disks from
becoming unrealistically cold at large radii and late times.
Our cloud irradiation serves as a stand-in for one neglected heat
source in clusters: irradiation from surrounding stars. This effect is
likely important for (a) very dense regions, and (b) late times when
disk radii stretch out to $10^4$ AU. Due to the wide variance in the
strength of this effect, we do not address heating by neighbors here.
There is also minor heating due to the accretion shock that feeds the
disk; however, KM06 have argued that this is negligible in general.

While our background heating is only important at late times, we do
not report results for $t>$2\,Myr as this may exceed the lifetime of
gas disks, even the low-mass ones \citep{2006ApJ...648.1206J}.  The
uncertainties in our procedure therefore have little effect on the
results we obtain.

We have now fully specified the conditions of thermal and mechanical
equilibrium for this disk, and we can use them to compute the
accretion rate. Equations (\ref{mdotloc}) and (\ref{Tdisk})
constitute two equations for the unknowns $Q$ and $\dot{M}_*$. For any
given $M_d$, $M_*$, and $\mathbf{J_d}$, we may solve them to determine
$\dot{M}_*$. This in turn also specifies the rate of change of the
disk mass
\begin{equation}
\dot{M}_d = \dot{M}_{\rm in} - \dot{M}_*.
\end{equation}
Note that $M_d$  can also be modified by disk fragmentation
and binary formation, as described in \S~\ref{fragmentation} and
\S~\ref{binaries}.

\subsection{An Outer Disk and the Braking Torque} \label{angmom}

When describing standard steady-state disks, one implicitly assumes
that when angular momentum is transported radially, it travels out to
large radii in an insignificant amount of mass. In our current model,
we effectively keep track of an ``inner" disk: the portion containing
the majority of the mass.
This justifies our choice of surface density profile $\Sigma\propto
r^{-1}$, since the radius at which this power-law slope is achieved is
also the radius that encloses most of the mass.
We allow for a small amount ($2\%$) of material raining in from the
core to be carried out with the angular momentum.

The disk's angular momentum is then equal to that of the infalling
material, in addition to the amount already in the disk, minus some
portion which has been transferred to this outer disk. The disk loses
a fraction $b_j$ of its angular momentum and a small amount of mass on
the viscous timescale $\tau_v = M_d/\Mdot_*$, {\em so long as matter
  is still accreting from the core}:
\begin{equation}
\label{j_disk}
\dot{\mathbf J}_d = {\mathbf j}_{\rm in} \dot{M}_{\rm in} - b_j
\left(\frac{\Mdot_{\rm in}}{\Mdot_*}\right)\frac{\Mdot_{\rm in}}{
  M_d}{\mathbf J}_d.
\end{equation}
As above, the subscript ``in" denotes newly accreted matter.  The factor
$(\Mdot_{\rm in}/\Mdot_*)$ is roughly unity in the main accretion
phase, but goes to zero when accretion stops.  We thus assume the
outer disk only applies a torque when it gains matter from the inner
disk.  Without accretion the outer disk has no effect, and thus after
accretion ends, the inner disk is free to expand self-similarly at
constant ${\mathbf J}_d$. We consider this a conservative approach,
considering that we do not treat effects like photoionization that
might remove material from the inner and outer disk, especially in
massive stellar clusters.

We consider $b_j = 0.5$ to be typical; in this case an accreting disk
loses about half its angular momentum each viscous time.  Since the
disk sheds mass at the same rate, this allows its radius to remain
comparable to the circularization radius of the infalling material.
Although our choice of $b_j$ is somewhat arbitrary we demonstrate that
our parametrization makes the disk evolution somewhat independent of
this value. See \S \ref{vary_fj} for discussion.
 
\subsection{Disk Fragmentation} \label{fragmentation}

Since we have now computed $\dot{M}_d$, $\dot{M}_*$, and
$\dot{\mathbf{J}}_d$, our model is almost complete. However, as
demonstrated by both previous analytic work (KM06) and numerical
simulations \citep{Krum2007a,Vorobyov06,LodRi05}, our
parameter space extends deeply into the regime where disk
fragmentation is expected. We must account for this to model disk
evolution. It is not our intent to follow the detailed evolution of
the fragments formed, nor their mass spectrum; we are interested
primarily in how they help the disk regulate $Q$.

In keeping with the approach outlined in \S~\ref{Approach}, we make
the important assumption that the disk fragments into small objects
when $Q$ drops below a critical value, $Q_{\rm{crit}}$, which we take
to be unity.  Other authors \citep{Gam2001, 2003MNRAS.339.1025R} have
pointed out the importance of a disk's thermal physics in setting the
fragmentation boundary.  In particular they find, in simulations with
imposed cooling, a critical value of $\tau_c \Omega$ above which disks
do not fragment, and below which they do.

Our fragmentation model reproduces these results (indeed, it is
calibrated to the same simulations) and we believe that the two views
are in fact equivalent.  Within our model, a disk whose $Q$ is close
to unity will be heated by accretion at a rate close to the critical
cooling rate found in these simulations.  In the absence of any
additional heating, the cooling rate must exceed the critical value in
order for $Q$ to fall below unity, so that fragmentation can commence.
In other words, since in our model $Q$ is calculated based on the
competition between cooling and the combination of viscous dissipation
and irradiation, if $Q$ falls below unity then it is
necessarily the case that the cooling rate is sufficient to overwhelm
viscous heating, and therefore to satisfy a cooling condition similar
to those identified by \citet{Gam2001} and
\citet{2003MNRAS.339.1025R}.

The benefit of our fragmentation model is that it can be easily
extended into the realistic regime of irradiated disks, whereas a
model that refers solely to the cooling time cannot.

We note, in support of our model, that we know no examples of disks
for which $Q<1$ that do not fragment, nor those with $Q>1$ that do.
Moreover, \cite{2003MNRAS.339.1025R} note that a sufficiently slowly
cooling disk reaches an equilibrium at a $Q$ value \emph{higher} than
unity; this is consistent with a heating rate that drops sharply
as $Q$ increases, as our accretion model would predict.

To implement fragmentation within our numerical models, we must
specify how much mass goes into fragments each time step when
$Q<1$. We first define a critical density, $\Sigma_{d,c}$:
\begin{equation} \label{sigma_crit}
\Sigma_{d,c} = \frac{c_s \Omega}{\pi G Q_{\rm{crit}}};
\end{equation}
a reduction of surface density from $\Sigma_d$ to $\Sigma_c$ would
return the disk to stability.
Because we expect fragmentation to happen over a dynamical time, we
assume that it depletes the disk surface density at the rate:
\begin{equation} \label{sigma_dot}
\dot{\Sigma}_{\rm frag} = -(\Sigma_d -\Sigma_{d,c}) \Omega,
\end{equation}
This rate is fast enough to ensure that $Q$ never dips appreciably
below $Q_{\rm{crit}}$.

For simplicity, we assume that while fragments contribute to the mass
of the disk, they do not enter in Toomre's stability parameter $Q$
except insofar as they contribute to the binding mass.  (One could
consider a composite $Q$: \citealt{2001MNRAS.323..445R}.)  Nor do we
follow the migration of fragments in the disk.  Instead, we allow them
to accrete onto the central star at the rate
\begin{equation} \label{mdotfrags}
\Mdot_{*,\rm frags} = \phi_f M_{\rm{frag}} \Omega,
\end{equation}
with $\phi_f = 0.05$. The assumption is simply that some fraction of
the fragments accrete each orbit.  Fragments form preferentially at
large distances from the star, and thus only a small amount of the
fragment mass will make it into the central star each orbit. Changing
this parameter by an order of magnitude only marginally alters the
disk evolution.

We also make the important assumption that disks will always fragment
to maintain stability, and allow accretion to proceed. While this is
likely a good assumption based on the existence of massive stars that
appear to have formed via disk accretion, the persistence of rapid
accretion during fragmentation has not been satisfactorily
demonstrated in numerical simulations.  See \S \ref{supersonic}.

\subsection{Binary Formation}\label{binaries}
A majority of stars, especially massive stars, are found in binary and
multiple systems. Though we present a very simplified scenario for
star formation, we do account for the possibility that a single
secondary star will form if $M_d> M_*$, that is, if the disk grows
unphysically large with respect to the central star.  (As we discussed
in \S~\ref{angmom}, this may well be conservative -- in the sense that
secondaries may form at even lower values of $\mu$, or at earlier
times through core fragmentation as described in \citealt{Bon2004}.)
When $\mu > 0.5$, we remove the excess mass and store it (and the
associated angular momentum) in a binary star. Because this tends to
happen before the disks have become very extended, we assume the
binary separation will be small; we therefore ignore the binary as a
source of angular momentum for the disk.  As with fragments, we assume
the disk is affected by the secondary star only through the increased
binding mass.  We make no attempt to account for its contribution to
the total luminosity.

\subsection{Summary of Model}\label{summary}

\begin{figure*}[]
\begin{center}
\includegraphics[scale=0.6]{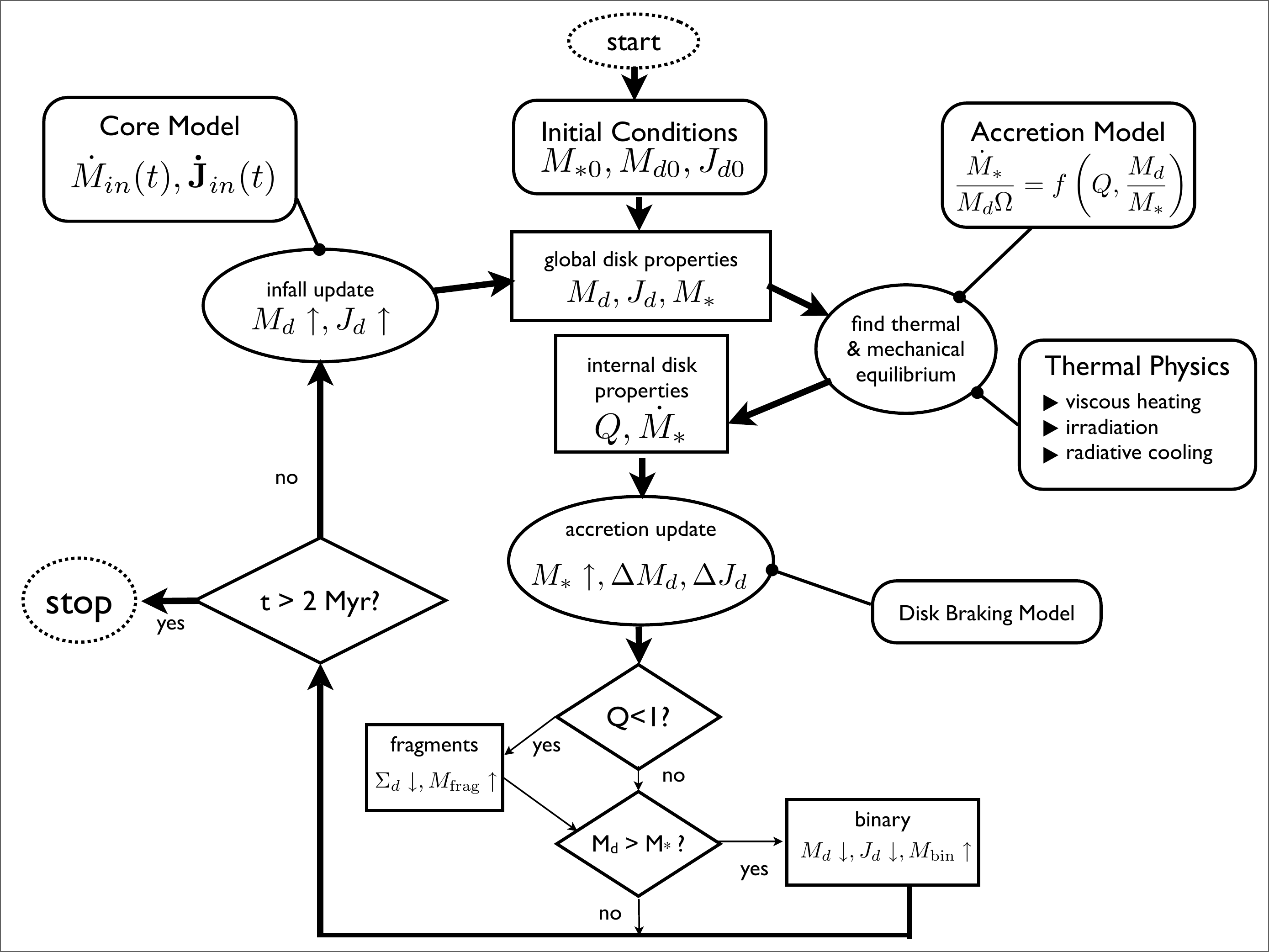}
\caption{{A simplified schematic of the decision tree in the code. The
    primitive variables, $M_d,M_*$, and $\mathbf{J_d}$, together with
    the core model, $\Mdot_{in}(t)$ and $\mathbf{\dot{J}}_{in}(t)$,
    allow for the determination of all disk parameters at each time
    step. Note that $c_s, Q$, and $\Mdot$ are solved for
    simultaneously. Once the self-consistent state is found, the
    values of $Q$ and $\mu$ determine whether either the binary or
    fragmentation regime has been reached. See \S \ref{summary} for a
    description of the elements in detail.}}
\label{flowchart}
\end{center}
\end{figure*}

We summarize our model via the flowchart shown in Figure
\ref{flowchart}, which illustrates a simplified version of the code's
decision tree. At a given time $t$ we know the current disk and star
mass, and the current angular momentum and mass infall rates as
prescribed in \S \ref{sfmodel} and \S \ref{j_acc}. We can calculate
$R_d$ and $\Sigma_d$ directly, and find the appropriate stellar
luminosity based on its evolution, current mass, and accretion
rate. Using these variables we self-consistently solve for the
appropriate temperature, $Q$, and disk accretion rate as described in
\S \ref{diskheat}. With this information in hand, we determine whether
the disk is stable, locally fragmenting, or forming a binary. If the
disk is stable, we proceed to the next iteration. If $Q<1$, then the
disk puts mass into fragments according to equation
[\ref{sigma_crit}]. If $\mu > 0.5$ we consider binary formation to
have occurred, and the net angular momentum and disk mass over the
critical threshold is placed into a binary (see \S \ref{binaries}). We
stop simulations after $2$ Myr for two reasons: first, the most
massive stars in our parameter space are significantly evolved and so
our stellar evolution models are no longer sufficient; and second,
because many other effects begin to dominate the disks appearance at
late stages due to gas-dust interaction and photo-evaporation
\citep{2007ApJ...666..976K}.

\section{Expected Trends} \label{expects}
Before examining the numerical evolution, it is useful to make a
couple analytical predictions for comparison.

First, can we constrain where disks ought to wander in the plane of
$Q$ and $\mu$?  This turns out to depend critically on the
dimensionless system accretion rate
\begin{equation} \label{Psi_in}
\Re_{\rm in} \equiv {\dot{M}_{\rm in} \over M_{*d} \Omega(R_{\rm
    circ})} = {\dot M_{\rm in} j_{\rm in}^3 \over G^2 M_{*d}^3}
\end{equation} 
which is the ratio of the mass accreted per radian of disk rotation
(at the circularization radius $R_{\rm circ}$) to the total system
mass $M_{*d} = M_* + M_d$.  Since the active inner disk has a radius
comparable to $R_{\rm circ}$, this controls how rapidly the disk gains
mass via infall.

The importance of $\Re_{\rm in}$ is apparent in the equation governing
the evolution of the disk mass ratio $\mu$:
\begin{eqnarray} \label{muEvoln} 
{\dot\mu\over \mu\Omega}& =& {\dot M_{\rm in}\over M_{*d}\Omega
}\left(\frac1\mu-1\right) - {\dot M_*\over M_d\Omega } \nonumber\\ &=&
{\Omega(R_{\rm circ})\over\Omega}\left(\frac1\mu-1\right) \Re_{\rm in}
- {\dot M_*\over M_d \Omega }.
\end{eqnarray} 
Since we consider $\dot M_*/(M_d\Omega)$ to be a function of $\mu$ and
$Q$, we must know the disk temperature to solve for $\mu(t)$.
Regardless, equation (\ref{muEvoln}) shows that larger values of
$\Re_{\rm in}$ tend to cause the disk mass to increase as a fraction
of the total mass.  We may therefore view $\Re_{\rm in}$ and $Q$ as
the two parameters that define disk evolution -- of which $\Re_{\rm
  in}$ is imposed externally and $Q$ is determined locally.

Moreover, $\Re_{\rm in}$ takes characteristic values in broad classes
of accretion flows, such as the turbulent core models we employ.
Suppose the rotational speed in the pre-collapse region is a fraction
$f_K$ of the Kepler speed, so that $j_{\rm in} = f_K r v_K(r) = f_K [G
  r M_c(r)]^{1/2}$, and suppose also that the mass accretion rate is a
fraction $\varepsilon f_{\rm acc}$ of the characteristic rate
$v_K(r)^3/G$.  Then,
\begin{equation}\label{PsiCharacteristic}
\Re_{\rm in} = {f_K^3 f_{\rm acc}\over \varepsilon^2} .
\end{equation} 
(In this expression, negative three powers of $\varepsilon$ appear
because the binding mass is $\varepsilon$ times smaller for the disk
than for the core; one of these is compensated by the reduction of the
accretion rate.)

In \S~\ref{sfmodel} we adopted the \cite{MT2003} model for massive
star formation due to core collapse of a singular, turbulent,
polytropic sphere in initial equilibrium.  Their equations (28), (35),
and (36) imply
\begin{equation}\label{f_acc} 
f_{\rm acc} = 0.84 (1-0.30\krho ) \left(3-\krho\over1+H_0\right)^{1/2}
\end{equation} 
within $2\%$, where $1+H_0\simeq 2$ represents the support due to
static magnetic fields \citep{1996ApJ...472..211L}.  (Note, their
equation [28] is a fit made by \citealt{2002Natur.416...59M} to the
results of \citealt{1997ApJ...476..750M}.)

KM06 predicted the turbulent angular momentum of these cores; our
parameter $f_K$ equals $(\theta_j\phi_j)^{1/2}$ in their paper.  Their
equations (25), (26), and (29) imply
\begin{equation}\label{f_K}
f_K = {0.49 \over \phi_B^{1/2}} {(1-\krho/2)^{0.42}\over
  (\krho-1)^{1/2} },
\end{equation}
with excursions upward by about $50\%$ and downward by about a factor
of three expected around this value; here $\phi_B\simeq 2.8$
represents the magnetic enhancement of the turbulent pressure. All
together, we predict
\begin{eqnarray}\label{PsiInEval}
\Re_{\rm in} &=& {0.10 \over \varepsilon^2 \phi_B^{3/2} }
\left(3-\krho\over1+H_0\right)^{1/2}{
  \left(1-\frac\krho2\right)^{1.26} \over (\krho-1)^{3/2}}
(1-0.30\krho) \nonumber \\ &\rightarrow& 0.02
\left(0.5\over\varepsilon\right)^2
\end{eqnarray} 
where the evaluation uses $1+H_0\rightarrow 2$,
$\phi_B\rightarrow2.8$, and $\krho\rightarrow 1.5$.

Importantly, $\Re_{\rm in}$ is a function of $(1+H_0)$, $\phi_B$,
$\epsilon$, and $\krho$, but {\em not} the core mass.  We therefore
expect similar values of $\Re_{\rm in}$ to describe all of present-day
massive star formation, at least insofar as these other parameters
take similar values.  Suppose, for instance, that the formation of
$10\Msun$ and 100$\Msun$ stars were both described by the same
$\Re_{\rm in}$.  According to equation (\ref{muEvoln}), the difference
in $\mu$ between these two systems would be controlled entirely by the
thermal effects that cause them to take different values of $Q$.

A few additional expectations regarding $Q$ itself can be gleaned from
the analytical work of \cite{ML2005} and KM06:
\begin{itemize}
\item[-] The Toomre parameter remains higher than unity for low-mass
  stars ($\lesssim 1 M_\odot$) in low-column cores ($\Sigma_{c,0}\ll
  1$), but falls to unity during accretion for massive stars and for
  low-mass stars in high-column cores;
\item[-] A given disk's $Q$ drops during accretion, reaching unity
  when the disk extends to radii beyond $\sim 150$\,AU (in the
  massive-star case), or to periods larger than $\sim$460\,yr (in the
  case of an optically thick disk accreting from a low-mass, thermal
  core).
\item[-] At the very high accretion rates characteristic of the
  formation of very massive stars ($\gtrsim
  1.7\times10^{-3}M_{\odot}$\,yr$^{-1}$), disk accretion is strongly
  destabilized by a sharp, temperature-dependent drop in the Rosseland
  opacity of dust.
\end{itemize}
(For more detailed conclusions, see the discussion surrounding
equation [35] in (KM06).)  With the help of equation (\ref{muEvoln})
we also deduce that more massive stars will have generally higher disk
mass fractions, because: $(1)$ they are described (in our model) by
the same value of $\Re_{\rm in}$; $(2)$ more massive stars achieve
lower values of $Q$; and $(3)$ in our model, lower $Q$ leads to lower
values of $\dot M_*/(M_d\Omega)$, so long as $Q>1.3$. The conclusion
that higher-mass stars have relatively more massive disks follows from
these three points by virtue of equation [\ref{muEvoln}].

More generally, {\em any} effect which causes $\dot M_*/(M_d\Omega)$
to drop (without affecting $\Re_{\rm in}$) will tend to increase
$\mu$, and vice versa; this conclusion is not limited to our adopted
disk accretion model.

Within our model, $\dot M_*/(M_d \Omega)$ increases with $Q$ unless
$1<Q<1.3$, in which case the dependence is reversed.  Disks ought
therefore to traverse from high $Q$ and low $\mu$, to low $Q$ and high
$\mu$, until $Q=1.3$; but for $1<Q<1.3$, $\mu$ and $Q$ should decline
together.  In physical terms, this reversal represents a flushing of
the disk due to the strong angular momentum transport induced by the
short wavelength gravitational instability.

We now turn to our suite of numerical models to test these
expectations.
\section{Results} \label{analysis}
We begin by examining the evolution of disks through their accretion
history for a range of stellar masses, determining when and if they
are globally or locally unstable, and the dominant mechanism for
matter and angular momentum transport through disk lifetimes. Next, we
explore how these results are influenced by varying the other main
physical parameters: $T_c$, $\Sigma_c$, $\alpha_{\rm{MRI}}$, and by
varying the angular momentum prescription.  For this purpose we first
define a fiducial sequence of models in \S\,\ref{fiducial}, and then
expand our discussion to the wider parameter space encompassed by the
aforementioned variables. Figures \ref{mdottracks} -- \ref{mu_q} show
results from our fiducial model, and Figures \ref{qvaries} and
\ref{muvaries} explore the effects of our environmental variables.

Because our prescription for disk accretion and fragmentation is
necessarily approximate, any specific predictions are unlikely to be
accurate.  We concentrate instead on drawing useful observational
predictions from our models' evolutionary trends.

\subsection{The fiducial model}\label{fiducial}
Our fiducial model explores a range of masses with a standard set of
parameters, which we list in Table \ref{fidpars}. For our exploration
of the stellar mass parameter space, we allow $\Sigma_c$ to vary as
$\Sigma_c = 10^{-1.84} ({M_c/ \Msun})^{0.75}$ (with an enforced minimum at 0.03 g cm$^-2$ so that $\Sigma_c$ varies from $0.03-1$g cm$^{-2}$ across the mass range
$0.5-120\Msun$). This relationship ensures that for our fiducial model, each system is forming at a
$\Sigma_c$ that is characteristic of observed cores. Enforcing this $\Sigma_c -M_{c}$ correspondence specifies the core radius.  We explore the effects of $\Sigma_c$ independently in \S \ref{sigvar}. All ``low mass" runs
that are shown e.g. $1\Msun$, have $\Sigma_{\rm{c,low}} = 0.03$ g
cm$^{-2}$, and ``high mass", e.g. $15\Msun$, have $
\Sigma_{\rm{c,high}} = 0.5$ g cm$^{-2}$. All systems start out with an
initial stellar mass of $0.10 \Msun$, disk mass of $0.01\Msun$ and
$j_d = 10^{19}$. Varying these parameters over an order of magnitude
effects the initial evolution for a few thousand years, but runs
converge quickly.  One can find pathological initial conditions,
particularly for small mass values. We believe this is due to the lack
of sensitivity of a one zone model. The initial disk radius is
calculated self-consistently from the amount of mass collapsed into
the system at the first time step, the initial $J_d$ is typically
smaller by a factor of a few than $j_{in}$.

\begin{table}
\begin{center}
\begin{tabular}{|c|c|c|}
\hline \hline Parameter & Fiducial & Range\\ \hline $b_j$ & 0.5 &
$0-1.0$\\ $\Sigma_{\rm {c,low}}$ & 0.03 g cm$^{-2}$& $0.03-1$ g
/cm$^2$\\ $\Sigma_{\rm{c,high}}$ & 0.5 g cm$^{-2}$ & $0.03-1$ g
/cm$^2$\\ $T_{c}$ & 20 K& $10-50$K \\ $\alpha_{\rm{MRI}}$ & 0.01 &
$0.001-0.1$\\ $\phi_f$ & 0.05 & $0 - 0.5$\\ $\varepsilon$ & 0.5&
N/A\\ \hline
\end{tabular}
\caption{Fiducial parameters for disk models for low and high mass
  stars, and the accompanying ranges explored. \label{fidpars}}
\end{center}
\end{table}%

\begin{figure*}
\begin{center}
\plotone{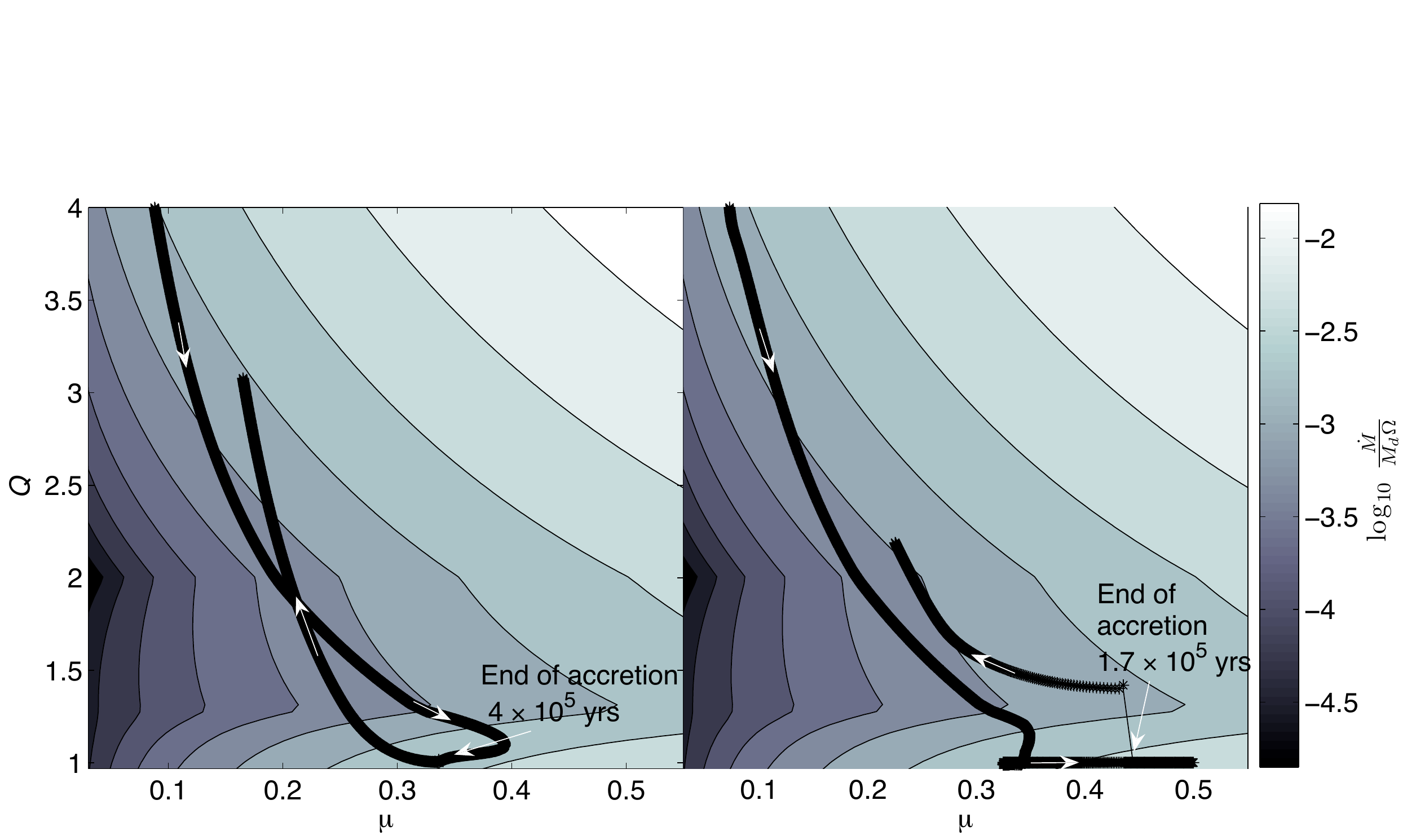}
\caption{{Evolutionary tracks in the $Q, \mu$ plane of a
    1$\Msun$(left) and 15 $\Msun$ (right) final star-disk system overlayed
    on the contours of our accretion model (contour spacing is
    identical to Figure \ref{mdotcontours}). The white arrows
    superposed on the tracks show the direction of evolution in
    time. The low mass star remains stable against fragmentation
    throughout its history, while the more massive star undergoes
    fragmentation and more violent variation in disk mass. The jump a the end of accretion in the 15 $\Msun$ system is due to the switch in the irradiation calculation. }}
\label{mdottracks}
\end{center}
\end{figure*}

\begin{figure}
\plotone{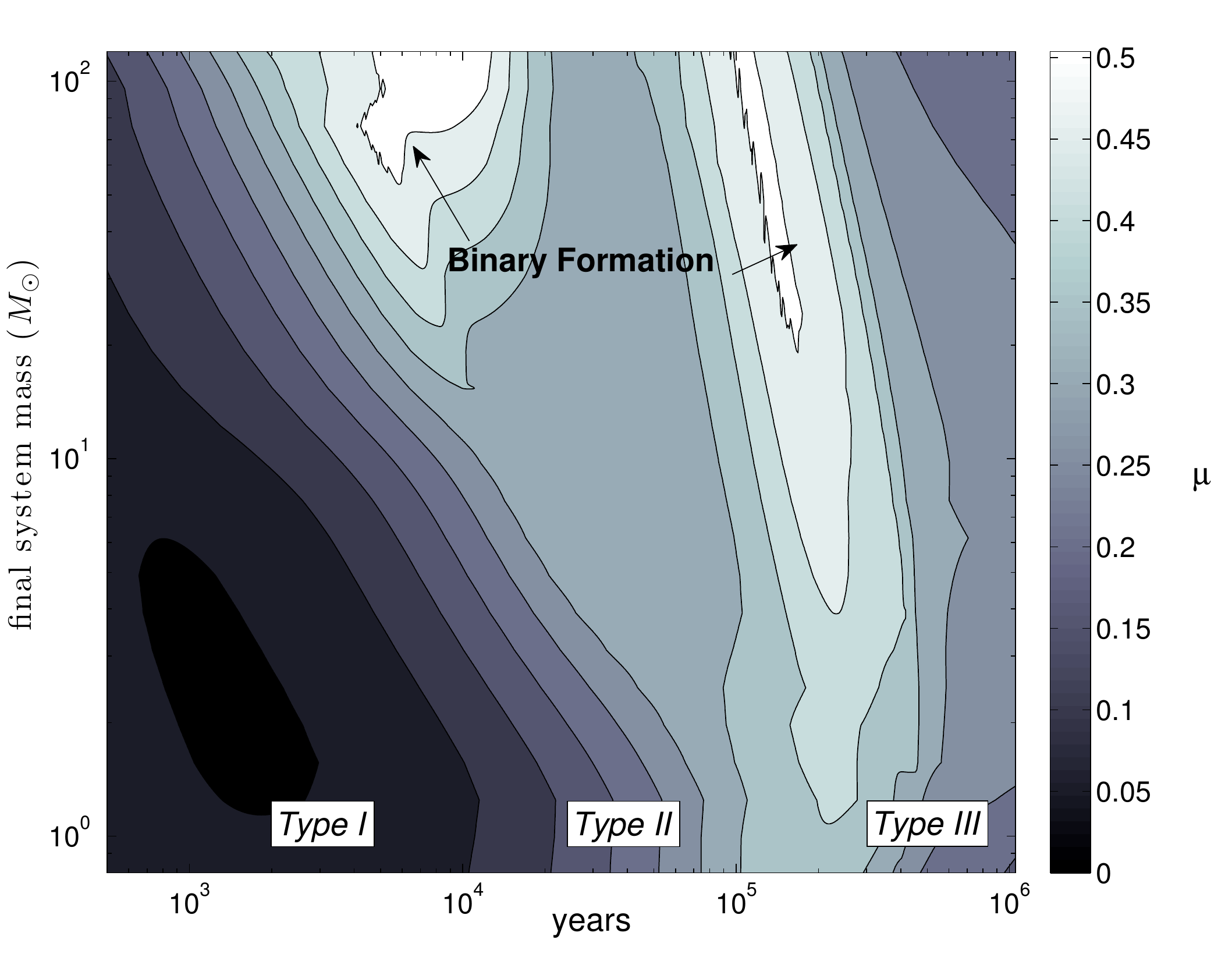}
\caption{{Contours showing the evolution of $\mu={M_d\over M_d + M_*}$
    for the fiducial sequence. Each contour shows an increase of 0.05
    in $\mu$. Again one can see the division into three regimes: low
    mass disks at early times, higher mass, unstable disks that may
    form binaries during peak accretion times, and low mass disks that
    drain following the cessation of infall. Systems destined to accrete up
    to $\sim 70 \Msun$ or more experience two epochs of binary
    formation in our model. In these systems the accretion from the
    core exceeds the maximum disk accretion rate very early, causing
    the disk mass to build up quickly.}}
\label{mu_q}
\end{figure}

As illustrated by the evolutionary tracks of accreting stars in the
$Q-\mu$ plane in Figure \ref{mdottracks}, our model agrees with the
general trends of previous work and with the expectations described in
\S\,\ref{expects}, in that low mass systems are stable and have low
values of $\mu$, while more massive systems undergo a period of strong
gravitational instability (\citealt[KM06]{Krum2007a}). Here we see
that as we go to higher stellar masses, disks spend more of their time
at high $\mu$ and undergoing disk fragmentation.  For stars of $\la
1\Msun$, $Q$ stays above unity, and the disks remain Toomre stable,
although still subject to gravitational instability due to their
non-negligable disk masses (see Figure \ref{mu_q}). (Note that due to
our abrupt shift in the disk irradiation model, there is a small
discontinuity in the temperature calculation at the end of accretion
which can cause unphysical fragmentation even at low masses, and a jump in Q at all masses.) The
expectation that $Q$ and $\mu$ evolve in opposite directions until
$Q<1.3$ is also roughly borne out. However, note that for the
$15\Msun$ star-disk system (right plot), the accretion rate is great enough that
there is a build up of mass in the disk once $Q$ reaches unity, and
the local instability saturates. This saturation leads to binary
formation (see \S \ref{form_binary}). 

\begin{figure}
\begin{center}
\plotone{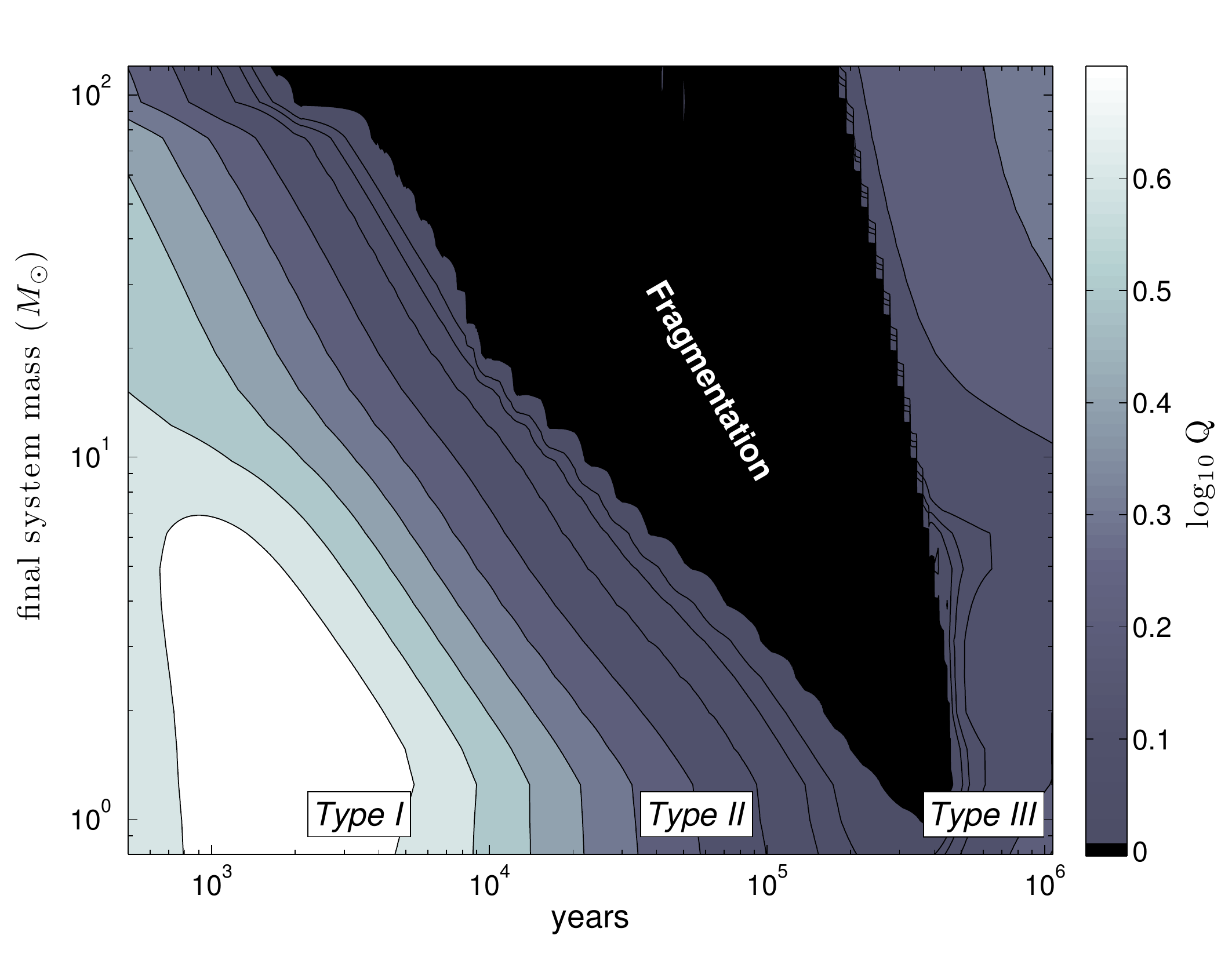}
\caption{{Contours of Q over the accretion history of a range of
    masses for the fiducial sequence. Masses listed on the y-axis are for the total star-plus-disk system final mass -- because the models halt at 2 Myr, some mass does remain in the disk.  Contours are spaced by  0.3 dex. At low final stellar masses, disks remain stable against
    the local instability throughout accretion. At higher masses, all
    undergo a phase of fragmentation. One can see three distinct
    phases in the evolution as described in \S \ref{analysis}. Disks
    start out stable, subsequently develop spiral structure as the
    disk mass grows and become unstable to fragmentation for sufficiently high
    masses. As accretion from the core halts, they drain onto the star
    and once again become stable.}}
\label{Qevolution}
\end{center}
\end{figure}

Figure \ref{Qevolution} shows the evolution of $Q$ through the
accretion history of a range of masses. We see that disks become
increasingly susceptible to fragmentation with increasing mass.  Disks
born from cores that are smaller than about $2\Msun$ remain stable
against fragmentation throughout their evolution, although we
expect moderate spiral structure \citep[as is seen in the models
  of][]{Lod04}.  Recall that with $\varepsilon=0.5$, a $2\Msun$ core
makes a $1\Msun$ star-disk system.  Figure \ref{mu_q} illustrates
the corresponding evolution of $\mu$ throughout the accretion history
for the same set of masses. As described in \S \ref{expects} the
typical disk mass increases with stellar mass. At high masses, binary
formation occurs during the peak of accretion just before $10^5$
years, and for stars $\ga 100\Msun$, there is an early epoch of binary
formation at roughly $10^4$ years.

We also see that for higher mass cores there are three relatively
distinct phases through which disks evolve:

\begin{enumerate}
\item[-]{Type I: Young, $<10^4$ yr systems, whose disks are described
  by small mass fractions and relatively high $Q$.  These would appear
  as early Class $0$ sources, deeply embedded in their natal clouds.}
\item[-]{Type II: Systems between $10^4-10^5$ yrs in age, whose disks
  are subject to spiral structure, and in high mass systems,
  fragmentation. Disk mass fractions are $\sim 30\%-40\%$, substantially
  higher than in Type I systems. These disks would appear in Class 0-I
  sources. }
\item[-]{Type III: Systems older than $10^5$ yrs, which have stopped
  accreting from the core, and consequently acquire low disk mass
  fractions as the disks drain away. These are the disks that are
  most like those observed in regions of LMSF as Class I
  objects.}
\end{enumerate}
These three stages serve as a useful prediction for future
observations; see \S \ref{observables} for more details.

\subsection{Influence of vector angular momentum}
\begin{figure}
\plotone{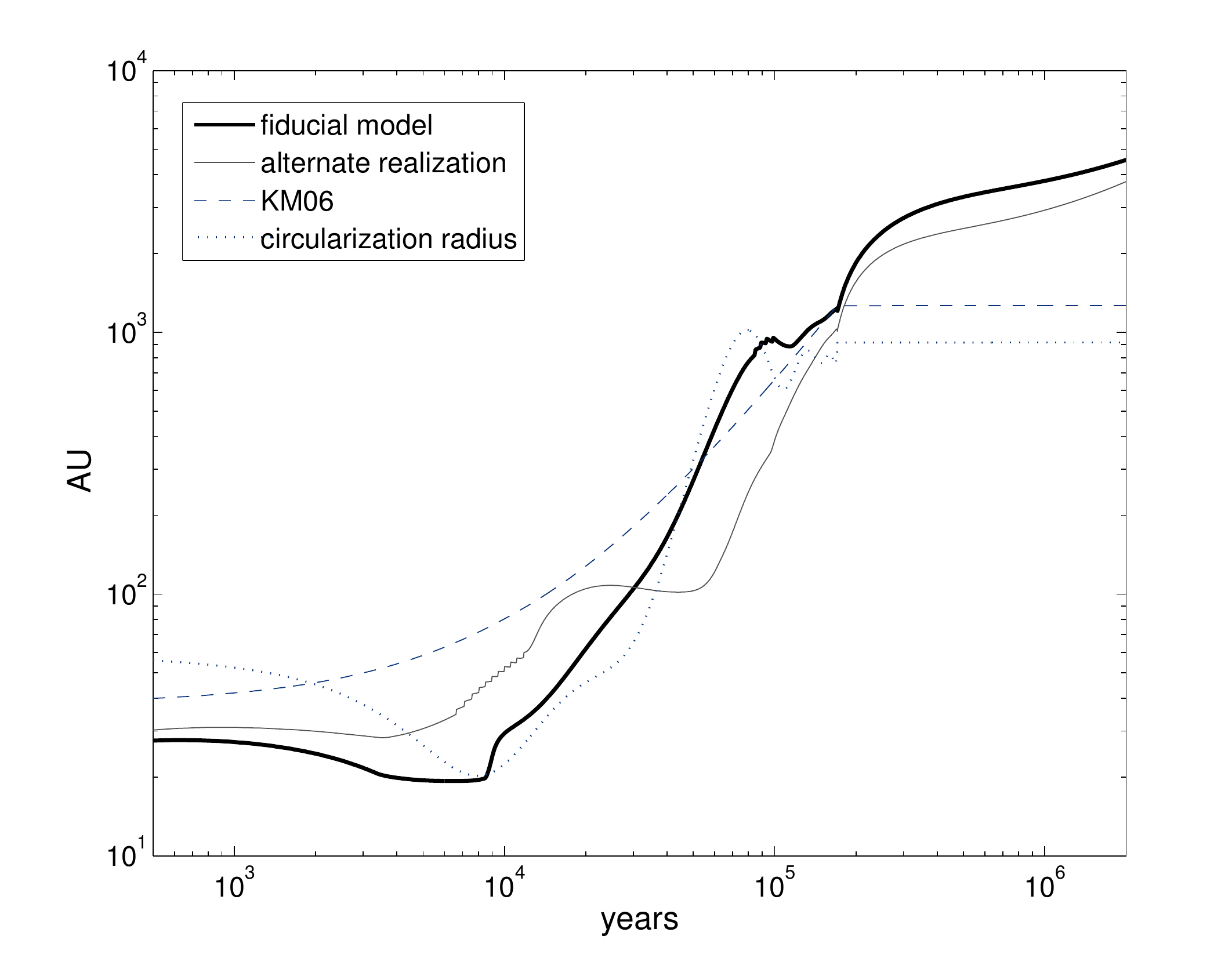}
\caption{Comparison of disk radius over the evolution of a 20 $\Msun$
  star-disk system in four cases: the KM06 analytic calculation, the
  circularization radius of the currently accreting material, and two
  realizations of the numerical model. The analytic case overestimates
  the expected radius at early times because it does not allow for
  cancellation of vector angular momentum. Similarly the
  circularization radius is an overestimate because the disk has no
  ``memory" of differently oriented j. At later times, the
  circularization radius approaches the standard radius calculation
  for that realization (thick black line) demonstrating the
  concentration of turbulent power at large scales. } \label{rad_comp}
\end{figure}

The accretion disk's radius plays a critical role in determining
whether or not the disk fragments. Consequently, we expect our results
to depend somewhat on effects that change the disk's specific angular
momentum. Because we track the vector angular momentum of the inner
disk, and because our turbulent velocity field is three dimensional,
we account for a possible misalignment between the disk's angular
momentum axis, $\hat \mathbf J$, and that of the infalling angular
momentum, $\hat \mathbf j_{\rm in}$.  The wandering and partial
cancellation that result provide a more realistic scenario than given
by the KM06 analytic approximations, in which vector cancellation is
accounted for only in an average sense. In practice, however, the disk
and infall remain aligned rather well (${\hat \mathbf J\cdot \hat
  \mathbf j_{\rm in}}\sim 0.8$), so misalignment plays only a minor
role in limiting the disk size.  This is illustrated by Figure
\ref{rad_comp}, in which we compare the disk radius in two numerical
realizations of the turbulent velocity field, against one in which
$j_{\rm in}$ has a fixed direction and obeys the KM06 formulae.  We
also plot the infall circularization radius $R_{\rm circ}$ (of one
numerical realization) for comparison. In general we find that the
analytic prescription slightly over-predicts the disk radius at early
times; this is partly due to ``cosmic'' variance in the numerical
realization, and partly due to disk-infall misalignment.

\subsection{Varying $\Sigma_{c}$} \label{sigvar}
We explore the effect of individual parameters by considering one or
two systems along our fiducial sequence, and varying parameters one by
one relative to their fiducial values.  First, we vary $\Sigma_c$ over
 0.03-1\,g\,cm$^{-3}$, spanning the range from isolated to intensely
clustered star formation \citep{1997ApJ...476..730P}. Column density affects
star formation in two primary ways: it influences the core radius (by
determining the confining pressure) and the accretion rate during
collapse (again, by setting the outer pressure and thus the velocity
dispersion). However, these two effects counteract one another:
smaller values of $\Sigma_c$ correspond to larger cores and larger,
thus more unstable disks ($R_{d} \propto \Sigma_c^{-1/2}$), but
smaller $\Sigma_c$ also leads to lower accretion rates and thus
stabler disks ($\Mdot \propto \Sigma_c^{3/4}$).  The thermal balance
of the disk midplane is affected by these trends.  An analysis by KM06
(see their equation [35]) shows that higher $\Sigma_c$ inhibits
fragmentation if the disk temperature is dominated by viscous heating
(which is proportional to the accretion rate), but enhances
fragmentation if irradiation dominates (when the increase in accretion
generated heating is insignificant), and that the two effects are
comparable along our fiducial model sequence.  We therefore expect
fragmentation to be quite insensitive to $\Sigma_c$, for massive star
formation along our fiducial sequence. This is precisely what we find
in our models: disks born from lower-$\Sigma_c$ cores, in lower
pressure environments, evolve in essentially the same way, but more
slowly.

In contrast, disks around low-mass stars -- those with final masses
comparable to the thermal Jeans mass -- are stable at low $\Sigma_c$
\citep[as predicted by][]{ML2005}, and
because irradiation dominates at larger radii, higher $\Sigma_c$ tend
to enhance fragmentation there. Figure \ref{qvaries} illustrates the
evolution of $Q$ for a $1\Msun$ accreting star for a range of column
densities.

\subsection{Varying $T_{c}$}
Observations of infrared dark clouds, and sub-mm core detections find
typical temperatures from $10-50$K \citep{2001ApJ...559..307J}. In our
models $T_c$ determines the amount of thermal, and therefore
turbulent, support: higher temperatures require less turbulent support
in the core.  Temperature also sets the thermal Jeans mass $M_{\rm
  th}$ within the \cite{MT2003} two component core model.  Accretion
from this thermal region leads to more stable disks; therefore, higher
core temperatures increase the mass of a star which can accrete
stably.

\begin{figure*}
\plotone{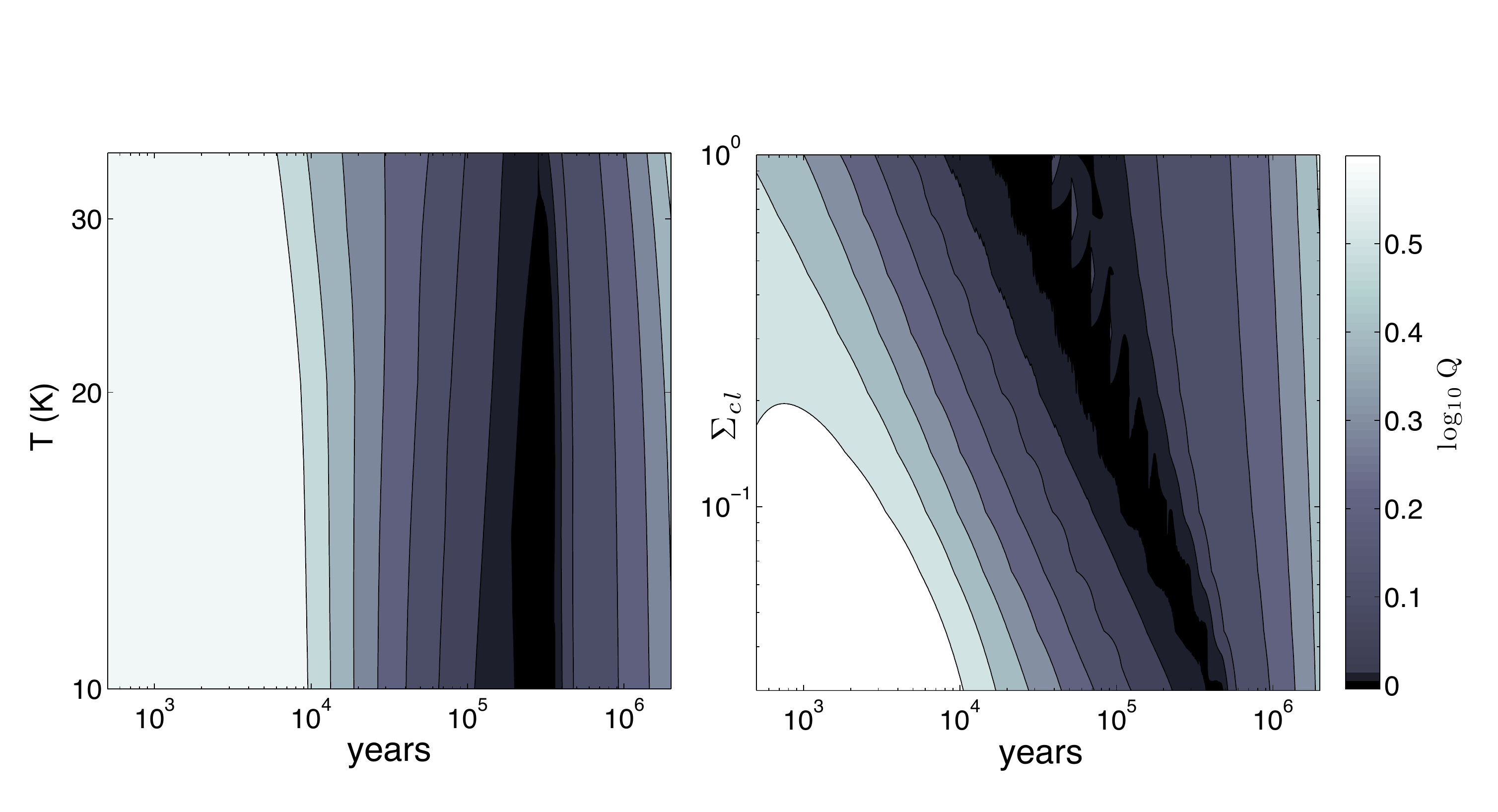}
\caption{Contours of Q showing the effect of initial core temperature $T_c$ (left)
  and  $\Sigma_c$  (right) on the evolution of a
  1 $\Msun$ final star-disk system. Contour spacing is 0.1
  dex (except the lowest two contours which are spaced by .05 dex). Increasing $\Sigma_c$ tends to marginally destabilize the disk,
  while higher temperatures stabilize the disk. We exclude temperatures too high for the $2\Msun$ core to collapse given its initial density, i.e., those above 40 K.}
\label{qvaries}
\end{figure*}

Figure \ref{qvaries} shows the evolution of $Q$ during the accretion
of a system with final mass $1\Msun$ over a range of temperatures (all
other parameters take their fiducial values). The difference in
evolution is negligible for high mass stars, as these accrete from
supersonic cores.

\subsection{Varying $\alphaMRI$} \label{MRI}
This work is not an exploration of the detailed behavior of the MRI;
we include it as the standard mechanism for accretion in the absence
of gravitational instabilities, which in most scenarios (aside perhaps
from low mass stars whose disks \citealt{2007arXiv0705.0421S} have argued may be strongly sub-Keplerian) overpower the MRI. However,
the strength of the MRI does influence the transition to
gravitationally dominated accretion in the $Q-\mu$ plane as shown in
Figure \ref{mdotcontours}.  The strength of the MRI also influences
the maximum disk mass obtained before gravitational instabilities set
in: higher values of $\alphaMRI$ reduce the influence of gravitational
instabilities by insuring that the disk drains more quickly, whereas
lower values expedite the transition to gravitational instability driven
accretion. Figure \ref{muvaries} shows the influence of $\alphaMRI$ on
$\mu$; the influence on $Q$ is less dramatic: the descent of $Q$
towards unity is marginally delayed for the strong MRI case.

\subsection{Varying $b_j$} \label{vary_fj}

\begin{figure}
\plotone{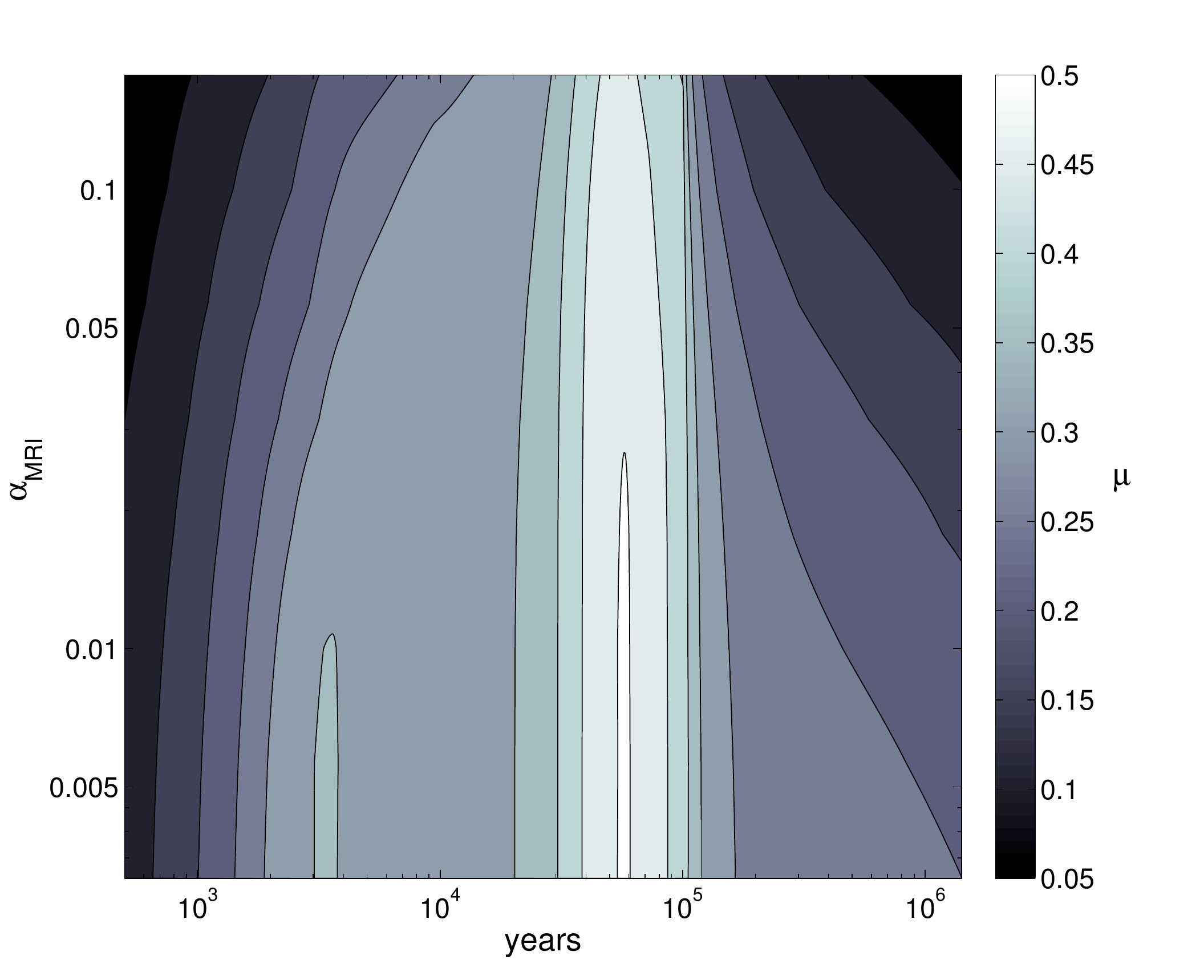} \plotone{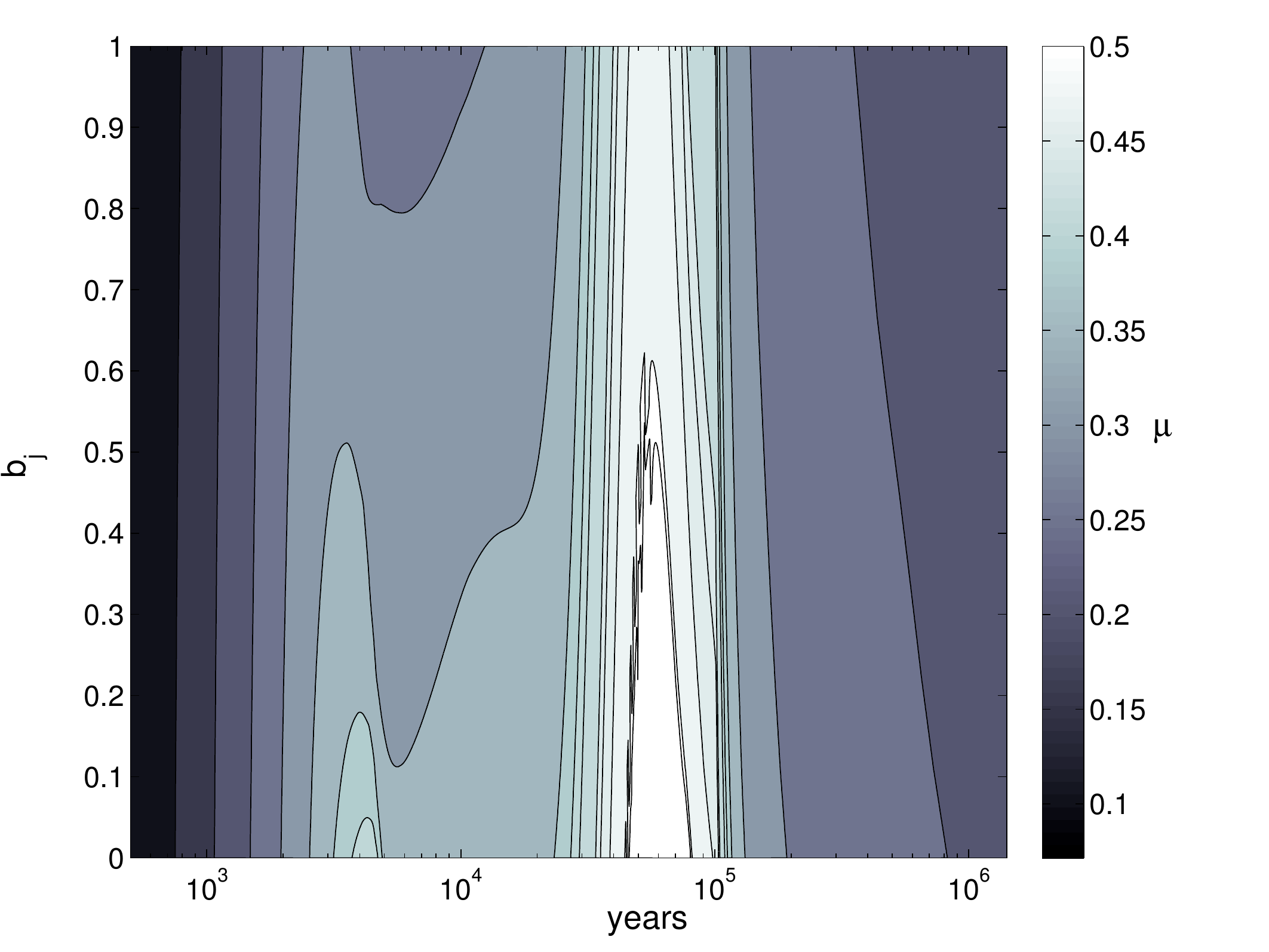}
\caption{Contours in $\mu$ illustrating the influence of varying
  $\alphaMRI$ (top) and the braking index, $b_j$ (bottom) for a star-disk system of final mass $15 \Msun$, the lowest mass at which a binary forms in our fiducial model. Contours of $\mu$ are spaced by 0.05. The upper plot shows the effect of varying $\alphaMRI$ from
  $10^{-2.5} - 10^{-1.5}$. While the change has little effect on the
  evolution of $Q$, the disk fraction $\mu$ decreases with increasing $\alphaMRI$. As a
  result, the mass at which binary formation begins is pushed to
  higher masses. The lower plot shows the effect of varying the braking index $b_j$. An increase of $b_j$ lowers the disk angular momentum, reducing the disk mass and inhibiting binary
  formation. Note that the variation in disk mass is only $\sim10\%$.}
\label{muvaries}
\end{figure}

Our most uncertain variable is the braking index $b_j$, which
determines the rate of angular momentum exchange with an outer disk.
However, disk evolution turns out to be rather insensitive to this
parameter. The primary reason for this is the concentration of power
in the turbulent velocity field on the largest scales: $j_{\rm in}$ is
always large compared to the disk-average $j$.  This reduces the
importance of the loss term in equation (\ref{j_disk}). As a result,
although the period in which the disk is fragmenting is reduced in the
high $b_j$ case, it is merely postponed by $\sim10^4$ yrs. The braking
index does have moderate influence on the disk mass during the peak of
accretion, and thus on the formation of binaries. Figure
\ref{muvaries} shows the evolution of $\mu$ for a system  accreting
towards $15\Msun$ from a $30\Msun$ core. Here one can see the
influence of $b_j$ on binary formation. Low values of $b_j$
corresponding to higher net angular momentum produce binaries at lower masses
by allowing the disk mass to grow larger. Notably, even for the
$15\Msun$ final mass star shown here, the smallest mass for which
binaries form in the fiducial model, the change in disk mass is only
$\sim 10\%$.

\subsection{The formation of binaries} \label{form_binary}
Within the context of our model for disk fission into a binary system,
(described in \S \ref{binaries}), the formation of a companion is
strongly dependent on the infalling angular momentum distribution, and
on the turbulent velocity profile of the particular core.  In our
fiducial model, binary formation occurs for cores above $30\Msun$. For
cores $\ga 140 \Msun$, there are two epochs of binary formation, the
first one at roughly $10^4$ years.  This mass boundary is quite
sensitive to our conservative threshold for disk fission, $\mu = 0.5$:
binaries may well form at lower values of $\mu$, and thus at lower
masses (see Figure \ref{mu_q}). The mass of the binaries that form
increases with initial core mass. This increase simply indicates that
the mass ratio exceeds the critical value for more time, as we do not
include a mechanism for accretion onto the binary. As such, we do not
predict values for the binary mass ratio $q$, but simply indicate the regimes in which
binary formation seems likely. The $30\Msun$ core cut-off is fairly
robust to variations in $\Sigma_{{c}}, T_{\rm{c}}$ and $b_j$ over the
ranges discussed above for our fiducial turbulent field. Cosmic
variance in the field from one realization to another has a much
larger effect on binary formation than any of our other model
parameters (aside from $\mu_{\rm{crit}}$).

In our fiducial model disk fission only occurs when the gravitational
instability has saturated and $Q \sim 1$. This means that the disk is
draining at the maximum rate given its mass. If matter is falling in
from the core more rapidly than this rate, the disk mass will
increase: if the accretion rate from the core exceeds the maximum rate
at $Q=1$ and $\mu=0.50$, disk fission occurs. In our fiducial model,
this corresponds to an accretion rate on to the disk:
${\Mdot_{\rm{in}} / {M_d \Omega}} = 10^{-2.36}$. The early epoch of
binary formation at very high masses is a consequence of this limit:
since the accretion rates begin to exceed the critical rate sooner,
the disk's mass increases earlier in its evolution. This critical
value is in agreement with the prediction of KM06 that disks are
sharply destabilized when accreting at rates higher than $1.7 \times
10^{-3} \Msun$ yr$^{-1}$.

 The time at which binaries form is also very dependent on the angular
 momentum profile. In the fiducial model, the lowest mass binaries
 form during the peak of accretion, at $\sim10^5$ years, but as
 the final system mass increases, binary formation pushes to earlier
 times $\sim 10^4$ years.  In certain runs we find earlier binary
 formation at smaller masses ($<10^4$ years) when there is a peak in
 the infalling angular momentum profile which rapidly sends $Q$
 towards unity. The presence of binaries in much of our parameter
 space illustrates that heavy circumbinary disks may be critical to
 binary evolution.

Observations suggest that a range of binary systems exist as a result
of variations in angular momentum as evidenced by the presence or lack
of disks around each component. Submillimeter observations of lower
mass objects in Taurus have revealed evidence for a binary with
circumstellar and circumbinary disks \citep{2003ApJ...586.1148O},
where the binaries are close enough to cause disk truncation ($\sim
45$ AU).  \cite{2004ApJ...605L.137A} have found another Class 0/I
binary system in NGC 1333 in which only the primary has a disk: the
diversity of systems is likely due to the variations in angular
momentum of the infalling material. As \cite{1997MNRAS.285...33B}
suggest, binaries forming from low angular momentum material will
likely not form their own disks, while those with higher angular
momentum may. It seems plausible that the absence or presence of
secondary disks is indicative of the formation process of the system.

As illustrated by these observations, the dependence we find on core
angular momentum is a sensible outcome: one expects the chance
rotation to have a stronger effect on multiplicity than other
parameters like temperature and density, which set the minimum
fragmentation mass. We emphasize that we are only exploring one
possible path for binary formation, and predict that disk
fragmentation is an important, if not the dominant mechanism at high
masses and column densities. This is especially true since, as argued by 
\cite{2006ApJ...641L..45K}, who has shown that for massive stars, once
the central core has turned on, the Jeans mass rapidly increases due
to the stellar luminosity, significantly reducing the possibility for
Jeans-instability induced core fragmentation.

\section{Observable Predictions} \label{observables}
Our models make strong predictions for the masses and morphologies of
disks during the embedded, accreting phase, and these will be directly
testable with future observations. Detailed calculations based on
radiation-hydrodynamic simulations of massive protostellar disks
indicate that disks with $\mu$ of a few tenths around stars with
masses $\ga 8$ $\Msun$, corresponding to embedded sources with
bolometric luminosities $\ga 10^4$ $L_{\odot}$, should produce levels
of molecular line emission that are detectable and resolvable with
ALMA in the sub-millimeter out to distances of a few kpc, and with the
EVLA at centimeter wavelengths at distances up to $\sim 0.5$ kpc
\citep{2007ApJ...665..478K}. The ALMA observations will be
particularly efficient at observing protostellar disks, since ALMA's
large collecting area will enable it to map a massive disk at high
resolution in a matter of hours. Dust continuum emission at similar
wavelengths should be detectable at considerably larger distances,
although the lack of kinematic information associated with such
observations makes interpretation more complex. Regardless of whether
dust or lines are used, observations using ALMA should be able to
observe a sample of hundreds of protostellar disks around embedded,
still-accreting sources, with masses up to several tens of $\Msun$.

The main observational prediction of our model is the existence of
type II disks -- those with $\mu$ of a few tenths or greater and $Q
\approx 1$ -- and the mass and time-dependence of the type II
phase. Examining Figures \ref{Qevolution} and \ref{mu_q}, we see that
our model predicts that protostellar cores with masses $\la 2$ $\Msun$
should experience only a very short type II disk phase, or none at
all. In contrast, cores with larger masses have type II disks for a
fraction of their total evolutionary time that gets larger and larger
as the core mass rises, reaching the point where type II disks are
present during essentially the entire class 0, accreting phase for
cores $\ga 100\Msun$ in mass.

Type II disks have several distinct features that should allow
observations to distinguish them from type I or type III disks, and
from older disks like those around T Tauri and Herbig AE stars. First,
since type II disks are subject to strong gravitational
instability, they should have strong spiral arms, with most of the
power in the $m=1$ or $m=2$ modes. This is perhaps the easiest feature
to pick out in surveys, since it simply requires observing the disk
morphology and can therefore be measured using continuum rather than
lines.

Second, because their self-gravity is significant, type II disks will
deviate from Keplerian rotation due to non-axisymmetric motions, and
will also be super-Keplerian in their outer parts compared to their
inner ones. The latter effect arises because, when the disk mass is
comparable to the stellar mass, the enclosed mass rises as one moves
outward in the disk. Recent work by \cite{2007ApJ...666L..37T}
provides a possible example of this phenomenon. The source HW2 in
Cepheus A is predicted to have a central mass of order $15\Msun$, and
a disk radius of $300$AU, with a temperature slightly under $200$ K
\citep{Pat2005, {2007ApJ...666L..37T}}. High resolution VLA observations
now show evidence of non-Keplerian rotation
\citep{2007ApJ...661L.187J}, consistent with our predictions for type
II disks.

Third, a type II disk is massive enough for the star-disk system
center of mass to be significantly outside of the stellar surface if
the disk possesses significant non-axisymmetry. As a result, the star
will orbit the center of mass of the system, and this will produce a
velocity offset of a few km s$^{-1}$ between the stellar velocity and
the zero velocity of the inner, Keplerian parts of the disk
\citep{1999A&A...350..694B,2003MNRAS.339.1025R,2007ApJ...665..478K}. This
should be detectable if the stellar velocity can be measured, which
may be possible using Doppler shifts of radio recombination lines for
stars producing hypercompact HII regions, or using proper motions for
stars with large non-thermal radio emission
\citep{2003ApJ...598.1140B}. In fact recent work by \cite{2007ApJ...666L..37T} has observed said offset. As suggested by
\citet{2001A&A...375..455L, 2003A&A...408.1015L}, one
could also look for the effect in the unresolved radio emission from
FU Orionis objects.

A final point concerns the limited range of the disk-to-system mass
ratio in our simulations, with $0.2<\mu<0.5$ during most of embedded
accretion (our type II disks).  The upper envelope of $\mu$ depends in
part on our binary fragmentation threshold $\mu=0.5$. However, in the
absence of disk fission, disks in our fiducial model never grow larger
than $\mu = 0.55$. The fact that most accretion occurs with $\mu\sim
0.3$ provides strong evidence that accretion disks do not become very
massive compared to the central point mass \citep[as argued
  by][]{{{1989ApJ...347..959A}}}.  Current observations such as those
of \cite{Ces2005} describe massive tori with sub-Keplerian rotation
and comparable infall and rotation velocities. These structures are
distinct from the disks that we model: our finding that disks hover
around $\mu = 0.3$ suggests that higher resolution observations may
reveal the Keplerian structures within the tori. The underlying
physical reason for this is that it is not possible to support a mass
comparable to the central star in a rotationally supported disk for
long periods of time; gravitational instabilities will destabilize
such a disk on orbital timescales, causing it to lose mass either
through rapid accretion or fragmentation.

\section{Conclusions} \label{conclusions}
We have constructed a simple, semi-analytic one-zone model to map out
the parameter space of disks in $Q - \mu$ space across a range of
stellar masses throughout the Class 0 and Class I stage, pushing into
the Class II phase. We include angular momentum transport driven by
two different mechanisms: gravitational instability and MRI transport modeled by
a constant $\alpha$. Our model for angular momentum infall is unique
in that we keep track of an inner and outer disk, and infall direction
so that cancellation may occur as the infall vector rotates.  We allow
for heating by the central star, viscous dissipation and a background
heat bath from the cloud accounting for both the optically thin and
optically thick limit within the disk and accreting envelope.  By
requiring that the disk maintain mechanical and thermal equilibrium,
we determine the midplane temperature at each time step, and thus $Q$
in the disk.

\subsection{Caveats}\label{supersonic}
In interpreting the results of our calculations, it is important to
keep several caveats in mind.  Our model for fragmentation, though
rooted in simulations, includes one
important assumption: no matter how violently unstable a disk
becomes, it can always fragment, return to a marginally stable state,
and continue accreting. While the existence of stars well into the mass
regime of fragmentation makes this outcome seem likely, it has yet to
be demonstrated in simulations.  Equally untested is the hypothesis
that when fragmentation is strong enough, i.e., when $\Mdot_{\rm in}
\gg c_s^3/G$ so that $Q\ll 1$ \citep{Gam2001}, accretion onto the
central star will be choked off.  KM06 have argued that accretion is
sharply destabilized when its rate exceeds $1.7\times10^{-3}\,
\Msun$\,yr$^{-1}$, due to a drop in the Rosseland opacity, and that
this may be related to the stellar upper mass limit.

In order to explore a wide parameter space, we do not carry out
detailed hydrodynamic calculations to determine the onset of
instability, but instead use results from previous simulations, and
develop analytic formulae that describe behavior intermediate between
the regimes which they explore. Although our approach is very
approximate, it can be made increasingly more realistic as additional
numerical simulations become available. Due to our one-zone
prescription, we cannot resolve spiral structure or measure the degree
of non-Keplerian motion. In addition, we do not follow the evolution
of fragments, nor their interaction with the disk. Although we allow
for the formation of binaries, we do not follow their evolution and
accretion, which limits our ability to make predictions about mass
ratios and angular momentum transfer between the disk and the
companion. Our model for angular momentum infall is responsible for
the largest uncertainty in our conclusions because different
realizations of the turbulent velocity field can alter the disk size
at a given epoch by a factor of a few. Nevertheless, these variations
are well within the analytic expectations for range of angular momenta
in cores (KM06). Moreover, our approach aims only to predict
characteristics of the outer accretion disk, and lacks the resolution
to track the radial profiles of the disk's properties.

Lastly, recall that our models rely on the fundamental assumption
(\S~\ref{Approach}) that a disk's behavior can be separated into
dynamical and thermal properties, and in particular that its dynamics
are governed primarily by its mass fraction $\mu$ and Toomre parameter
$Q$.

With these caveats in mind, we summarize our results for two different
regimes: $<2\Msun$ and $>2\Msun$.

\subsection{The Low Mass Regime} \label{lowmasscon}
Our fiducial models predict that low mass stars will have higher
values of $\mu$ than typically assumed during early phases of
formation. However, they should remain stable against fragmentation
throughout their evolution, dominated by
MRI, long wavelength gravitational instability, and once again MRI through their evolution
through the three types of disks discussed in \S
\ref{analysis}. During the main accretion phase, disks will have
masses of order $30\%$ of the system mass. Typical outer radii are of
order 50 AU, with outer temperatures of $40$ K during the main
accretion phase, dropping to $\sim 10$ K at 2 Myr. The surface density
is $10-20$ g cm$^{-2}$ during the main accretion phase, dropping off
rapidly at late times causing the disk to become optically thin to its
own radiation.  As accretion shuts down, and disks grow due to conservation of angular
momentum, two key effects must be considered: truncation and heating
by other stars. At distances of $~1000$ AU, very tenuous disks are
prone to truncation by passing stars particularly in denser clusters
where average stellar densities are as high as $10^5$ stars pc$^{-3}$
\citep{1998ApJ...492..540H}. Similarly, as the disk edge extends
towards other, potentially more luminous stars, the actual flux
received will increase, heating the disk above the $\sim10$K
temperature that we routinely find \citep{2006ApJ...641..504A}.

For core column densities more typical of high-mass star forming
regions, local instabilities do set in, despite the stabilizing
influence higher temperatures associated with these regions
(neglecting the effects of nearby stars). This implies that
environment may be important in understanding disk evolution.

In contrast to our previous work \citep{KM06,ML2005}, we find fragmentation at smaller radii. This is primarily due to our modified model for $\alpha_{\rm GI}$, which predicts lower accretion rates and consequently more fragmentation then previously assumed. We note that our results for low-mass systems (final mass $\sim 1 \Msun$) are rather sensitive to details of the model, such as the value of $\alpha_{ \rm MRI}$ and the way it is combined with $\alpha_{\rm GI}$.

\subsection{The High Mass Regime} \label{highmasscon}
For more massive stars, we find high values of $\mu \sim 0.35$ and an
extended period of local fragmentation as the accretion rates
peak. Temperatures at the disk outer edge at $\sim 200$ AU approach
100K for systems $>15\Msun$ during accretion. surface densities hover
around 50 g cm$^{-2}$ during the main accretion phase, although by 2
Myr, the disks become optically thin in the FIR, as expected. Binary
formation occurs regularly for cores of order $30\Msun$ and higher,
though as discussed in \S \ref{binaries} this is strongly dependent on
the cosmic variance of the angular momentum: cores as small as
$20\Msun$ form binaries in our model when there is excess angular
momentum infall. Although fragments accrete with the disk according to
equation [\ref{mdotfrags}], more massive stars maintain a small mass
in fragments ($10^{-1}-10^{-2} \Msun$) in the disk when we end our
simulations, suggesting that fragments may persist to form low mass
companions, as predicted in KM06 and suggested by the simulations of
\cite{Krum2007a}.

Unlike their low mass counterparts, the conclusions we draw for
massive stars are minimally effected by the environmental variables in
our model. For the entire range of temperatures, densities, and nearly
all angular momentum realizations, the conclusions listed above hold
true.

\acknowledgments The authors are grateful to an anonymous referee for insightful comments that improved and clarified this work. The authors would also like to thank Norm Murray, Debra
Shepherd, Guiseppe Lodato, and W. Ken Rice for helpful discussions. KMK is
supported by a U. Toronto fellowship. CDM's research is funded by
NSERC and the Canada Research Chairs program. MRK is supported through
Hubble Fellowship grant \#HSF-HF-01186 awarded by the Space Telescope
Science Institute, which is operated by the Association of
Universities for Research in Astronomy, Inc., for NASA, under contract
NAS 5-26555. This research was supported in part by the National Science Foundation under
Grant No. PHY05-51164.


\end{document}